\documentclass[preprint,12pt]{elsarticle}
\pdfoutput=1
\usepackage[utf8]{inputenc}
\usepackage{graphicx}
\usepackage{amssymb}
\usepackage{amsmath}
\usepackage{color}
\usepackage{todonotes}
\usepackage{listings}
\usepackage{hyperref}
\usepackage{verbatim}
\usepackage{verbatimbox}
\usepackage{soul}
\usepackage{subcaption}
\usepackage{fancyvrb}
\usepackage[english]{babel}
\usepackage{braket}
\usepackage{appendix}
\usepackage{float}
\usepackage{natbib}

\usepackage[version=3]{mhchem}
\graphicspath{ {./images/} }

\newcounter{bla}

\makeatletter
\newcommand*\bigcdot{\mathpalette\bigcdot@{.5}}
\newcommand*\bigcdot@[2]{\mathbin{\vcenter{\hbox{\scalebox{#2}{$\m@th#1\bullet$}}}}}
\makeatother

\definecolor{codegreen}{rgb}{0,0.6,0}
\definecolor{codegray}{rgb}{0.5,0.5,0.5}
\definecolor{codepurple}{rgb}{0.58,0,0.82}
\definecolor{backcolour}{rgb}{0.95,0.95,0.92}
 
\lstdefinestyle{mystyle}{
    backgroundcolor=\color{backcolour},   
    commentstyle=\color{codegreen},
    keywordstyle=\color{magenta},
    numberstyle=\tiny\color{codegray},
    stringstyle=\color{codepurple},
    basicstyle=\footnotesize,
    breakatwhitespace=false,         
    breaklines=true,                 
    captionpos=b,                    
    keepspaces=true,                 
    numbers=none,                    
    numbersep=5pt,                  
    showspaces=false,                
    showstringspaces=false,
    showtabs=false,                  
    tabsize=2
}
\usepackage{todonotes}

\journal{Computer Physics Communications}

\begin{document}

\begin{frontmatter}

\title{{\sc MechElastic}: A Python Library for Analysis of Mechanical and Elastic Properties of Bulk and 2D Materials}

\author[a,b]{Sobhit Singh\corref{author}}
\author[a]{Logan Lang}
\author[a]{Viviana Dovale-Farelo}
\author[a]{Uthpala Herath}
\author[a]{Pedram Tavadze}
\author[c]{François-Xavier Coudert}
\author[a]{and Aldo H. Romero\corref{author}}

\cortext[author] {Corresponding authors.\\\textit{E-mail address:} sobhit.singh@rutgers.edu; alromero@mail.wvu.edu}
\address[a]{Department of Physics and Astronomy, West Virginia University, Morgantown, WV 26505-6315, USA}
\address[b]{Department of Physics and Astronomy, Rutgers University, Piscataway, NJ 08854, USA}
\address[c]{Chimie ParisTech, PSL University, CNRS, Institut de Recherche de Chimie Paris, 75005, Paris, France}

\begin{abstract}

The {\sc MechElastic} Python package evaluates the mechanical and elastic properties of bulk and 2D materials using the elastic coefficient matrix ($C_{ij}$) obtained from any {\it ab-initio} density-functional theory (DFT) code. The current version of this package reads the output of VASP, ABINIT, and Quantum Espresso codes (but it can be easily generalized to any other DFT code) and performs the appropriate post-processing of elastic constants as per the requirement of the user. This program can also detect the input structure's crystal symmetry and test the mechanical stability of all crystal classes using the Born-Huang criteria. Various useful material-specific properties such as elastic moduli, longitudinal and transverse elastic wave velocities, Debye temperature, elastic anisotropy, 2D layer modulus, hardness, Pugh's ratio, Cauchy's pressure, Kleinman's parameter, and Lame's coefficients, can be estimated using this program. Another existing feature of this program is to employ the ELATE package [\href{https://iopscience.iop.org/article/10.1088/0953-8984/28/27/275201/meta}{J. Phys.: Condens. Matter  28, 275201 (2016)}] and plot the spatial variation of several elastic properties such as Poisson's ratio, linear compressibility, shear modulus, and Young's modulus in three dimensions. Further, the {\sc MechElastic} package can plot the equation of state (EOS) curves for energy and pressure for a variety of EOS models such as Murnaghan, Birch, Birch-Murnaghan, and Vinet, by reading the inputted energy/pressure {\it versus} volume data obtained {\it via} numerical calculations or experiments. This package is particularly useful for the high-throughput analysis of elastic and mechanical properties of materials.

\end{abstract}


\end{frontmatter}
{\bf PROGRAM SUMMARY}

\begin{small}
\noindent
{\em Program Title:} {\sc MechElastic} \\
{\em Licensing provisions:} GPLv3\\
{\em Programming language:} Python\\
{\em Supplementary material:}\\
{\em Nature of problem:} To automatize and simplify the analysis of elastic and mechanical properties for bulk and 2D materials, especially for high-throughput DFT calculations. \\
{\em Solution method:} This Python program addresses the above problem by parsing the elastic coefficient matrix obtained from the first-principles DFT calculations, does the appropriate post-processing to evaluate the elastic properties, and performs the mechanical stability tests for all crystal classes using the generalized Born-Huang criteria. 
It can also be used to carry out the equation of state analysis to study the structural phase transitions. 
{\sc MechElastic} can be downloaded using the below link:\\ \href{https://github.com/romerogroup/MechElastic}{https://github.com/romerogroup/MechElastic} \\
\textbf{Installation} ({\it via} PyPI): pip install mechelastic, or pip3 install mechelastic 

\end{small}

\section{Introduction}
\label{sec:intro}

The elastic and mechanical properties of any material are among the key properties that must be thoroughly investigated to facilitate the proper integration of that material into the emerging technology. One way to analyze the elastic and mechanical response of materials is {\it via} the  first-principles density-functional theory (DFT) calculations.
The success of DFT~\cite{HK1964, KS1965, hafner_wolverton_ceder_2006, CURTAROLO2005} not only enabled us to predict the properties of materials from {\it ab-intio} calculations, but it also expedited the discovery and design of novel materials with a given set of properties for desired practical applications. 
Modern DFT codes, $e.g.,$ ABINIT~\cite{gonze2002first, gonze2009abinit, gonze2016recent, aldo2020}, VASP~\cite{Kresse1996, Kresse1999}, SIESTA~\cite{Soler_2002}, Quantum Espresso~\cite{QE-2017, QE-2009}, exciting~\cite{Gulans_2014}, WIEN2k~\cite{WIEN2k2020}, CASTEP~\cite{CASTEP}, ELK~\cite{dewhurst2016elk}, and CRYSTAL~\cite{PERGER20091753, Dovesi2018}, offer the capability to efficiently compute the elastic stiffness tensor of materials to a remarkable accuracy (\textit{i.e.}, to the accuracy of the employed DFT method). In fact, such calculations have been performed in recent years and materials databases containing the DFT calculated elastic tensor of hundreds of bulk and two-dimensional materials have been 
developed~\cite{Setyawan2011, AJAIN2011, Jain2013, deJong2015, COUDERT2016, Choudhary2017, ChoudharyPRB2018}. 
Further, various software tools~\cite{YU2010} such as {\sc ELATE}~\cite{elate, elate2016}, AELAS~\cite{AELAS2017},  ElaStic~\cite{ElaStic2013}, and {\sc ELAM}~\cite{MARMIER2010}, to the best of our knowledge, have been developed to carry out the analysis of elastic properties from first-principles calculations. 

With the ever-increasing power of computing resources and burgeoning interest in the high-throughput DFT calculations~\cite{Morgan2004, Munter_2009, SetyawanCMP2010, Curtarolo2013}, there is a growing demand to develop software tools that can facilitate the automation of the elastic and mechanical analysis of bulk as well as of 2D materials, from the knowledge of the DFT computed elastic tensor. Moreover, access to the full elastic tensor enables one to determine numerous other important elastic, mechanical, and thermodynamic properties of materials vital for the screening of materials in the process of materials discovery and design~\cite{nye1985, mavko2020, deJong2015, Siwar2019}. Therefore, it is of crucial interest to have a software tool that can calculate and/or estimate the material-specific properties obtainable from the elastic tensor and be employed in a high-throughput manner. 

Considering the above requirements, we develop the {\sc MechElastic} package, a robust open-source Python library, that can facilitate the analysis of the elastic and mechanical properties of bulk and 2D materials in a manual way as well as in a high-throughput manner. The {\sc MechElastic} package reads the DFT calculated elastic tensor data obtained from various DFT codes (current implementation supports the output  generated from the VASP~\cite{Kresse1996, Kresse1999}, ABINIT~\cite{gonze2002first,gonze2009abinit,gonze2016recent, aldo2020}, 
and Quantum Espresso~\cite{QE-2017, QE-2009} codes, but it can be easily generalized to any other DFT code), stores the elastic tensor following the Voigt notation~\cite{nye1985}, and carries out the calculation of several important elastic moduli and mechanical parameters as per user's requirement. This package can also perform the Born-Huang mechanical stability tests~\cite{bornhuang1954} for all crystal classes.  Furthermore, elastic wave velocities, hardness, Debye temperature, and melting temperatures can be estimated using this package. One can also use this package to plot the equation of state (EOS) curves and investigate the structural phase transitions in the spirit of the Murnaghan, Birch, Birch-Murnaghan, and Vinet EOS models~\cite{birch1938effect, birch1947finite, murnaghan1951finite, birch1952elasticity, birch1977isotherms, vinet1989universal, stacey2001finite}.

Notably, we have interfaced the {\sc MechElastic} package with the {\sc ELATE} software~\cite{elate2016}. This interface enables one to analyze the three-dimensional spatial distribution of linear compressibility, shear modulus, Young's modulus, and Poisson's ratio, and make the desired plots offline, \textit{i.e.}, without requiring access to the website of {\sc ELATE}~\cite{elate}. This feature is particularly useful for characterization of the elastic anisotropy and auxetic features in materials~\cite{Siwar2019,Hearmon1946, OrtizPRL2012, Coudert2013, COUDERT2016, Lakesreview2017, Dagdelen2017, Pavlic2017, SinghPRB2018, singh2018structural}. 
It is worth mentioning that our package offers several capabilities, as described below, that are not yet offered by other existing software tools such as {\sc ELATE}~\cite{elate, elate2016}, AELAS~\cite{AELAS2017},  ElaStic~\cite{ElaStic2013}, and {\sc ELAM}~\cite{MARMIER2010}. 

This paper focuses on the presentation of the {\sc MechElastic} package, details the underlying algorithms and numerical implementations, and provides essential instructions to users. 
The outline of this paper is as follows: 
In Section~\ref{sec:elastictensor}, we briefly explain some basic but necessary details of the elastic tensor and introduce the Voigt notation for crystalline systems. 
In Section~\ref{sec:mechanical}, we present the Born-Huang mechanical stability test criteria for different crystal classes and discuss the role of crystal symmetries on the elastic tensor. Section~\ref{sec:moduli} contains the details of the methods and mathematical expressions employed for evaluation of elastic, mechanical, and thermodynamic properties. 
Section~\ref{sec:hardness} describes the methods used for the estimation of hardness and presents a guide to choose the best hardness model for a given crystal class and type (insulator/semiconductor/metal) of material. 
Section~\ref{sec:2d} focuses on the elastic properties and mechanical stability of 2D materials. 
Section~\ref{sec:elate} summarizes the integration details of the ELATE~\cite{elate} software tool into the {\sc MechElastic} package. 
Section~\ref{sec:eos} is dedicated to various implemented EOS models. 
In Section~\ref{sec:algorithm}, we present an overview of the {\sc MechElastic} library, and 
finally, in Section~\ref{sec:features}, we present some examples illustrating the capabilities of the {\sc MechElastic} package, 
which is being followed by summary in Section~\ref{sec:summary}.

\section{Elastic Tensor}
\label{sec:elastictensor}
Under the homogeneous deformation of crystal, the generalized form of stress-strain Hooke's law reads~\cite{nye1985} 

\begin{equation}
    \sigma_{ij} = C_{ijkl} ~ \epsilon_{kl},
    \label{eq:cij}
\end{equation}

where $\sigma_{ij}$ and $\epsilon_{kl}$ are the homogeneous two-rank stress and strain tensors, respectively. $C_{ijkl}$ denotes the fourth-rank elastic stiffness tensor.
Equation~\ref{eq:cij} represents a set of nine equations having a total of 81 $C_{ijkl}$ coefficients in a compact form. The symmetry of the stress and strain tensors implies 

\begin{equation}
    \sigma_{ij} = \sigma_{ji} \implies   C_{ijkl} =C_{jikl}, \text{~and} \\ 
    \label{eq:ij_symm}
\end{equation}
\begin{equation}
    \epsilon_{kl} = \epsilon_{lk} \implies   C_{ijkl} =C_{ijlk}. 
    \label{eq:kl_symm}
\end{equation}

These relations reduce the total number of independent $C_{ijkl}$ coefficients from 81 to 36. 
The stress–strain relation can be derived from a strain energy density-functional (U) through the work-conjugate relation~\cite{Leyu2016}, as given below
\begin{equation}
    \sigma_{ij} = \frac{\partial U}{\partial \epsilon_{ij}}. \\ 
    \label{eq:energy}
\end{equation}

Equation~\ref{eq:cij} suggests 
\begin{equation}
    C_{ijkl} = \frac{\partial \sigma_{ij}}{\partial \epsilon_{kl}}  \implies  C_{ijkl} = \frac{\partial^{2}U }{\partial \epsilon_{kl} \partial \epsilon_{ij}}.  \\ 
\end{equation}
Here, we note that $C_{ijkl} = C_{klij}$ due to the arbitrariness of the order of differentiation. These so-called major symmetries of the stiffness tensor further reduce the total number of independent $C_{ijkl}$ coefficients from 36 to 21.  

Using the matrix notation~\cite{nye1985}, we can abbreviate the four-suffixes stiffness tensor $C_{ijkl}$ into a two-suffixes stiffness tensor, \textit{i.e.}, the first two suffixes are abbreviated into a single suffix ranging from 1 to 6, and the last two suffixes are  abbreviated in the same way~\cite{nye1985}. 
To do this, we utilize the symmetry of the second-rank stress and strain tensors and write them as six-dimensional vectors in an orthonormal coordinate system. For instance,

\begin{equation}
    \sigma_{ij} = 
    \begin{bmatrix}
  \sigma_{xx} & \sigma_{xy} & \sigma_{xz} \\
  \sigma_{yx} & \sigma_{yy} & \sigma_{yz} \\
  \sigma_{zx} & \sigma_{zy} & \sigma_{zz}
    \end{bmatrix}
\\ 
    \label{eq:sigma}
\end{equation}
is simplified to a six-dimensional vector in the Voigt notation as: \\
$(\sigma_{xx}, \sigma_{yy}, \sigma_{zz},
  \sigma_{yz},\sigma_{xz},\sigma_{xy}) \equiv (\sigma_1, \sigma_2, \sigma_3, \sigma_4, \sigma_5, \sigma_6)$. \\

Similarly, $\epsilon_{ij}$ can be simplified to: \\
$(\epsilon_{xx}, \epsilon_{yy}, \epsilon_{zz},
  \epsilon_{yz},\epsilon_{xz},\epsilon_{xy}) \equiv (\epsilon_1, \epsilon_2, \epsilon_3, \epsilon_4, \epsilon_5, \epsilon_6)$. \\ 

Thus, in the Voigt notation the stress-strain relationship can be expressed as follows 
\begin{equation}
    \begin{bmatrix}
  \sigma_{1} \\
  \sigma_{2} \\
  \sigma_{3} \\
  \sigma_{4} \\
  \sigma_{5} \\
  \sigma_{6}
    \end{bmatrix} = 
    \begin{bmatrix}
C_{11}  &  C_{12} &   C_{13}&    C_{14} &   C_{15} & C_{16}\\
C_{21}  &  C_{22} &   C_{23}&    C_{24} &   C_{25} & C_{26}\\
C_{31}  &  C_{32} &   C_{33}&    C_{34} &   C_{35} & C_{36}\\
C_{41}  &  C_{42} &   C_{43}&    C_{44} &   C_{45} & C_{46}\\
C_{51}  &  C_{52} &   C_{53}&    C_{54} &   C_{55} & C_{56}\\
C_{61}  &  C_{62} &   C_{63}&    C_{64} &   C_{65} & C_{66}
      \end{bmatrix}
  \begin{bmatrix}
  \epsilon_{1} \\
  \epsilon_{2} \\
  \epsilon_{3} \\
  \epsilon_{4} \\
  \epsilon_{5} \\
  \epsilon_{6}
      \end{bmatrix}.
    \label{eq:voigt_stiff}
\end{equation}

Similarly, the elastic compliance tensor ($s_{ij} = C_{ij}^{-1}$) can be written as  
\begin{equation}
    \begin{bmatrix}
  \epsilon_{1} \\
  \epsilon_{2} \\
  \epsilon_{3} \\
  \epsilon_{4} \\
  \epsilon_{5} \\
  \epsilon_{6}
    \end{bmatrix} = 
    \begin{bmatrix}
s_{11} &   s_{12} &   s_{13}  &  s_{14} &   s_{15} & s_{16}\\
s_{21} &   s_{22} &   s_{23}  &  s_{24} &   s_{25} & s_{26}\\
s_{31} &   s_{32} &   s_{33}  &  s_{34} &   s_{35} & s_{36}\\
s_{41} &   s_{42} &   s_{43}  &  s_{44} &   s_{45} & s_{46}\\
s_{51} &   s_{52} &   s_{53}  &  s_{54} &   s_{55} & s_{56}\\
s_{61} &   s_{62} &   s_{63}  &  s_{64} &   s_{65} & s_{66}
      \end{bmatrix}
  \begin{bmatrix}
  \sigma_{1} \\
  \sigma_{2} \\
  \sigma_{3} \\
  \sigma_{4} \\
  \sigma_{5} \\
  \sigma_{6}
      \end{bmatrix}.
    \label{eq:voigt_compl}
\end{equation}

The abbreviations in the Voigt notation are done according to the following scheme: 
$11 \rightarrow 1$, $22 \rightarrow 2$, $33 \rightarrow 3$, $23,32 \rightarrow 4$, $31,13 \rightarrow 5$, and $12,21 \rightarrow 6$. Here $1, 2, 3, 4, 5,$ and $6$ numerals denote $xx, yy, zz, yz, zx,$ and $xy$ components, respectively, of the two-rank stress and strain tensors in an orthonormal coordinate system (note that $yz = zy$, $zx = xz$, and $xy = yx$). Henceforth, we describe the elastic stiffness tensor within the Voigt notation, {\it i.e.}, by a symmetric $6 \times 6$ matrix as defined in Eq.~\ref{eq:voigt_stiff}. 

Considering the major symmetries of the stiffness tensor, as discussed above, and the crystal symmetries, the total number of independent coefficients can be reduced to 3, 5, 7, 7, 9, and 13 for cubic, hexagonal, rhombohedral, tetragonal, orthorhombic, and monoclinic crystal systems, respectively~\cite{nye1985, MouhatPRB2014}. For triclinic systems, all 21 $C_{ij}$ coefficients remain independent. 


Expert users may choose to skip the next Section (Section~\ref{sec:mechanical}) and move on to Section~\ref{sec:moduli} for information regarding the evaluation of various parameters and elastic moduli obtainable from the $C_{ij}$ tensor.

\section{Mechanical Stability Criteria and Role of Crystal Symmetries}
\label{sec:mechanical}

The knowledge of $C_{ij}$ matrix enables one to test the mechanical (or elastic) stability of an unstressed crystal using the well-known Born-Huang mechanical stability criteria~\cite{born1940, bornhuang1954, MouhatPRB2014, WuPRB2007}. A crystal can be considered mechanically stable if and only if $C_{ij}$ matrix is positive definite. Although it is a necessary condition, it is not sufficient for generic class of crystal systems. Mouhat and Coudert~\cite{MouhatPRB2014} have elegantly described the necessary and sufficient conditions to test the mechanical stability for all crystal classes. 
We refer the reader to Ref.~\cite{MouhatPRB2014} for a detailed information of the total number of independent $C_{ij}$ coefficients and mechanical stability conditions for all eleven Laue crystal classes. 

In the following, we present the form of the symmetric $C_{ij}$ matrix along with the mechanical stability conditions for various crystal classes that we employ in the {\sc MechElastic} package. We use symbols $\bullet$ and $\cdot$ to represent the nonzero and zero components of the symmetric elastic tensor, respectively. Space group (SPG) numbers are also listed for each crystal class. The following information is collected from Refs.~\cite{nye1985, bornhuang1954,  MouhatPRB2014, WuPRB2007, Singh_PRM2018}. 

\subsection{Cubic: SPG\#195--230}
\vspace{0.15cm}

Cubic crystal systems have only three independent $C_{ij}$ coefficients, which are $C_{11}$, $C_{12}$, and $C_{44}$. 
\vspace{0.4cm}

\begin{equation}
\begin{bmatrix}
 C_{11}   &  C_{12}    &  C_{12}   &    \cdot   &   \cdot  &   \cdot  \\
 \bullet  &  C_{11}    &  C_{12}   &   \cdot   &   \cdot  &  \cdot  \\
 \bullet  &   \bullet  &  C_{11}   &   \cdot   &   \cdot  &  \cdot  \\
          &            &             &   C_{44}   &  \cdot   &  \cdot  \\
          &             &            &            &   C_{44}   &  \cdot  \\
          &             &            &            &            &  C_{44} 
\end{bmatrix}
\label{cubic_cij}
\end{equation}
\vspace{0.4cm}

The mechanical stability conditions for cubic systems are:
\begin{equation}
C_{11} - C_{12} > 0,~~ 
C_{11} + 2C_{12} > 0,~~\text{and}~
C_{44} > 0. \\ 
\label{eq:cubic_stability}
\end{equation}

\subsection{Hexagonal: SPG\#168--194}
\vspace{0.15cm}

Hexagonal crystal systems have five independent $C_{ij}$ coefficients that are $C_{11}$, $C_{12}$, $C_{13}$, $C_{33}$, and $C_{44}$. 
\vspace{0.4cm}

\begin{equation}
\begin{bmatrix}
 C_{11}   &   C_{12}   &   C_{13}    &   \cdot   &   \cdot  &  \cdot  \\
 \bullet  &   C_{11}   &   C_{13}    &   \cdot   &   \cdot  &  \cdot  \\
 \bullet  &   \bullet  &   C_{33}    &   \cdot   &   \cdot  &  \cdot  \\
          &            &             &   C_{44}  &   \cdot  &  \cdot  \\
          &             &            &           &  C_{44}  &  \cdot  \\
          &             &            &           &          &  (C_{11}-C_{12})/2 
\end{bmatrix}
\label{hexa_cij}
\end{equation}
\vspace{0.4cm}

The mechanical stability conditions for this system are:
\begin{equation}
C_{11} - C_{12} > 0,~~ 
2C_{13}^{2} < C_{33} (C_{11} + C_{12}),~~\text{and}~
C_{44} > 0. \\ 
\label{eq:hex_stability}
\end{equation}

\subsection{Rhombohedral class: SPG\#143--167}
\vspace{0.15cm}

Rhombohedral crystals can be divided into two classes: I (Laue class $\Bar{3}m$), and II (Laue class $\Bar{3}$)~\cite{MouhatPRB2014}. Rhombohedral class II has seven independent $C_{ij}$ coefficients ($C_{11}$, $C_{12}$, $C_{13}$, $C_{14}$, $C_{15}$, $C_{33}$, and $C_{44}$), as shown below. One coefficient, $C_{15}$, vanishes in the rhombohedral class I, thus leaving only six nonzero independent coefficients in the class I. 
\vspace{0.4cm}

\begin{equation}
\begin{bmatrix}
 C_{11}   &   C_{12}   &   C_{13}   &    C_{14}  &   C_{15}   &  \cdot  \\
 \bullet  &   C_{11}   &   C_{13}   &  - C_{14}  & - C_{15}   &  \cdot  \\
 \bullet  &   \bullet  &   C_{33}   &   \cdot    &   \cdot    &  \cdot  \\
 \bullet  &  \bullet   &            &   C_{44}   &  \cdot     & - C_{15}   \\
 \bullet  &  \bullet   &            &            &   C_{44}   &  C_{14}  \\
          &            &            &   \bullet  &  \bullet   & (C_{11} - C_{12})/2
\end{bmatrix}
\label{rhom_cij}
\end{equation}
\vspace{0.4cm}

The generic mechanical stability conditions for rhombohedral class I and II are: 
\begin{equation}
\begin{split}
C_{11} - C_{12} > 0,~~ C_{44} > 0,~~ \\ 
C_{13}^{2} < \frac{1}{2}C_{33} (C_{11} + C_{12}),~~\text{and}~\\
C_{14}^{2} + C_{15}^{2} < \frac{1}{2}C_{44} (C_{11} - C_{12}) \equiv C_{44}C_{66}. \\
\end{split}
\label{eq:rhom_stability}
\end{equation}

\subsection{Tetragonal: SPG\#75--142}
\vspace{0.15cm}

Similar to the rhombohedral crystal systems, tetragonal crystal systems can be divided into two classes: I (Laue class $4/mmm$) and II (Laue class $4/m$)~\cite{MouhatPRB2014}. Class I has six independent $C_{ij}$ coefficients ($C_{11}$, $C_{12}$, $C_{13}$, $C_{33}$, $C_{44}$, and $C_{66}$), whereas class II has seven independent $C_{ij}$ coefficients with $C_{16} \neq 0$, as shown below. 
\vspace{0.4cm}

\begin{equation}
\begin{bmatrix}
 C_{11}   &   C_{12}   &   C_{13}    &    \cdot   &   \cdot  &   C_{16}  \\
 \bullet  &   C_{11}   &   C_{13}     &   \cdot   &   \cdot  & - C_{16}  \\
 \bullet  &   \bullet  &    C_{33}  &   \cdot   &   \cdot  &  \cdot  \\
          &            &             &   C_{44}   &  \cdot   &  \cdot  \\
          &             &            &            &   C_{44}   &  \cdot  \\
 \bullet  &  \bullet    &            &            &            &  C_{66} 
\end{bmatrix}
\label{tetra_cij}
\end{equation}
\vspace{0.4cm}

The generic mechanical stability conditions for both tetragonal classes are:
\begin{equation}
\begin{split}
C_{11} - C_{12} > 0,~~ 
2C_{13}^{2} < C_{33} (C_{11} + C_{12}),~~ \\ 
C_{44} > 0,~~\text{and~~} 2C_{16}^{2} < C_{66} (C_{11} - C_{12}). \\ 
\end{split}
\label{eq:tetr_stability}
\end{equation}

\subsection{Orthorhombic class: SPG\#16--74}
\vspace{0.15cm}

Orthorhombic crystal systems have nine independent $C_{ij}$ coefficients as shown below.

\vspace{0.4cm}

\begin{equation}
\begin{bmatrix}
 C_{11}   &   C_{12}   &   C_{13}    &   \cdot      &   \cdot   &   \cdot  \\
 \bullet  &   C_{22}   &   C_{23}     &  \cdot      &   \cdot   &   \cdot  \\
 \bullet  &   \bullet  &    C_{33}  &   \cdot    &   \cdot  &  \cdot  \\
          &           &              &   C_{44}   &  \cdot   &   \cdot   \\
          &            &            &            &   C_{55}   &  \cdot   \\
          &            &            &            &            &  C_{66} 
\end{bmatrix}
\label{orth_cij}
\end{equation}
\vspace{0.4cm}

The generic mechanical stability conditions for  orthorhombic class are: 
\begin{equation}
\begin{split}
C_{11} > 0,~~ C_{11}C_{22} > C_{12}^{2},  \\
C_{44} > 0,~~C_{55} > 0, ~~C_{66} > 0, \\
C_{11}C_{22}C_{33} +2C_{12}C_{13}C_{23} - C_{11}C_{23}^{2} - C_{22}C_{13}^{2} - C_{33}C_{12}^{2} > 0. \\ 
\end{split}
\label{eq:orth_stability}
\end{equation}

\subsection{Monoclinic class: SPG\#3--15}
\vspace{0.15cm}

Monoclinic crystal systems have thirteen independent $C_{ij}$ coefficients as shown below.

\vspace{0.4cm}

\begin{equation}
\begin{bmatrix}
 C_{11}   &   C_{12}  &   C_{13}    &   \cdot      &   C_{15}   &   \cdot  \\
 \bullet  &   C_{22}   &   C_{23}     &  \cdot      &   C_{25}  &   \cdot  \\
 \bullet  &  \bullet   &    C_{33}  &   \cdot    &   C_{35}  &  \cdot  \\
          &           &              &   C_{44}   &  \cdot   &  C_{46}   \\
  \bullet   &   \bullet   &    \bullet     &            &   C_{55}   &  \cdot   \\
          &            &            &    \bullet   &            &  C_{66} 
\end{bmatrix}
\label{mono_cij}
\end{equation}
\vspace{0.4cm}

The generic mechanical stability conditions for the standard monoclinic class are~\cite{WuPRB2007}: \\ 
$C_{ii} > 0, ~~\forall~ i\in \{1, 6\}, $ \\ 
$C_{11} + C_{22} + C_{33} + 2(C_{12} + C_{13} + C_{23}) > 0, $ \\
$C_{33}C_{55} - C_{35}^{2} > 0, $ \\
$C_{44}C_{66} - C_{46}^{2} > 0, $ \\ 
$C_{22} + C_{33} - 2C_{23} > 0, $\\
$C_{22}(C_{33}C_{55} - C_{35}^{2}) + 2C_{23}C_{25}C_{35} - C_{23}^{2}C_{55} - C_{25}^{2}C_{33} > 0, $~\text{and} \\
$2[C_{15}C_{25}(C_{33}C_{12} - C_{13}C_{23}) + C_{15}C_{35}(C_{22}C_{13} - C_{12}C_{23}) + C_{25}C_{35}(C_{11}C_{23} - C_{12}C_{13})] - [C_{15}^{2}(C_{22}C_{33} - C_{23}^{2}) + C_{25}^{2}(C_{11}C_{33} - C_{13}^{2}) + C_{35}^{2}(C_{11}C_{22} - C_{12}^{2})] + C_{55}g > 0, $\\ 
\begin{equation}
\begin{split} 
\text{where, } g = C_{11}C_{22}C_{33} - C_{11}C_{23}^{2} - C_{22}C_{13}^{2} - 
C_{33}C_{12}^{2} + 2C_{12}C_{13}C_{23}.
\end{split}
\label{eq:mono_stability}
\end{equation}


\subsection{Triclinic class: SPG\#1--2}
\vspace{0.15cm}

Triclinic crystal systems can have all 21 nonzero independent $C_{ij}$ coefficients.

\vspace{0.4cm}

\begin{equation}
\begin{bmatrix}
 C_{11}   &   C_{12}  &   C_{13}    &   C_{14}     &   C_{15}   &    C_{16}  \\
          &   C_{22}   &   C_{23}     &   C_{24}      &   C_{25}  &    C_{26}  \\
          &            &    C_{33}  &    C_{34}    &   C_{35}  &   C_{36}  \\
          &           &              &   C_{44}   &   C_{45}   &  C_{46}   \\
           &             &                &            &   C_{55}   &   C_{56}   \\
          &            &            &             &            &  C_{66} 
\end{bmatrix}
\label{tric_cij}
\end{equation}
\vspace{0.4cm}

The generic expressions for the mechanical stability conditions for triclinic class are quite complex~\cite{MouhatPRB2014}. One can check the positive definiteness of the $C_{ij}$ matrix to ensure the mechanical stability of triclinic systems~\cite{MouhatPRB2014}.

\section{Elastic and Mechanical Properties of Materials}
\label{sec:moduli}

In this section, we present the formulae used for the evaluation of various parameters and moduli related to the elastic and mechanical properties of materials. 

Ideally, the stress and strain tensors (or fields) vary in position space $\rm{\bf{r}}$ within a polycrystalline material. Hence, the generic expression for Hooke's law in a homogeneous linear elastic polycrystal (in a simplified Voigt notation) is  
\begin{equation}
\sigma_{i}({\bf{r}}) = C_{ij}({\bf{r}})~\epsilon_{j}({\bf{r}}).  
\label{eq:gen_hookes}
\end{equation}

In a polycrystalline material, the evaluation of effective elastic constants and mechanical properties requires an averaging scheme (in $\rm{\bf{r}}$ space) to determine the average stress $\langle \sigma\rangle$ and strain $\langle \epsilon \rangle$ under the approximation of a ``statistically uniform" sample, where 

\begin{equation}
\langle \sigma\rangle = \frac{1}{V} \int \sigma({\bf{r}}) d{\bf{r}}, ~~~\text{and}~~~ \\
\langle \epsilon \rangle = \frac{1}{V} \int \epsilon({\bf{r}}) d{\bf{r}}. 
\label{eq:avg}
\end{equation}
Here, $V$ is the volume. In Equation~\ref{eq:avg} we assume that $\sigma({\bf{r}})$ and $\epsilon({\bf{r}})$ vary slowly and continuously in the position space of a polycrystal. 

Voigt (1928)~\cite{voigt1928} and Reuss (1929)~\cite{Reuss1929} proposed two different averaging schemes to obtain the effective elastic constants.
Voigt's scheme, also known as Voigt bound, relies on the  assumption that the strain field is uniform throughout the sample, \textit{i.e.}, $\epsilon(\rm{\bf r})$ is independent of ${\bf{r}}$. Whereas, Reuss's scheme, also known as Reuss bound, relies on the  assumption that the stress field is uniform throughout the sample, \textit{i.e.}, $\sigma(\rm{\bf r})$ is independent of ${\bf{r}}$. The Voigt and Reuss bounds provide an upper and lower estimate of the actual elastic moduli, respectively. 

In 1952, Hill~\cite{Hill1952} noted that an arithmetic mean of the Voigt and Reuss bounds, also known as the Voigt-Reuss-Hill (VRH) average, provides a good estimate of the elastic moduli, which is often close to the experimental data~\cite{mavko2020}. Here it is worth mentioning that, in addition to the aforementioned three bounds, numerous other bounds have been reported in literature, which are nicely summarized by Kube and de Jong in Ref.~\cite{KubeMaarten2016}. 

In the {\sc MechElastic} package, we employ the Voigt~\cite{voigt1928} (subscript $V$), Reuss~\cite{Reuss1929} (subscript $R$), and VRH~\cite{Hill1952} (subscript $VRH$) averaging schemes to obtain an estimate of the effective elastic moduli. The expressions of various elastic moduli for all three averaging schemes are given below~\cite{mavko2020}. \\

{\bf Voigt averaging scheme (upper bound):} 
\begin{equation}
    \text{Bulk modulus},~~ 
    K_{V} = \frac{1}{9} [\left(C_{11}+C_{22}+C_{33}\right) + 2\left(C_{12}+C_{23}+C_{31}\right)].
    \label{eq:Kv}
\end{equation}

\begin{equation}
\begin{aligned}
    \text{Shear modulus},~~ 
    G_{V} = \frac{1}{15} [\left(C_{11}+C_{22}+C_{33}\right)-\left(C_{12}+C_{23}+C_{31}\right) \\  + 3 \left(C_{44}+C_{55}+C_{66}\right)].
    \end{aligned}
    \label{eq:Gv}
\end{equation}

{\bf Reuss averaging scheme (lower bound):} 
In the Reuss averaging scheme~\cite{Reuss1929} we use the elastic compliance tensor $s_{ij}$ to compute the elastic moduli. 

\begin{equation}
    \text{Bulk modulus},~~ 
    \frac{1}{K_{R}} = \left(s_{11}+s_{22}+s_{33}\right) + 2\left(s_{12}+s_{23}+s_{31}\right).
    \label{eq:Kr}
\end{equation}

\begin{equation}
\begin{aligned}
    \text{Shear modulus},~~ 
   \frac{15}{G_{R}} = 4 \left(s_{11}+s_{22}+s_{33}\right)-4 \left(s_{12}+s_{23}+s_{31}\right) \\ + 3 \left(s_{44}+s_{55}+s_{66}\right). 
   \end{aligned}
    \label{eq:Gr}
\end{equation}

{\bf Voigt-Reuss-Hill (VRH) average:} 
In this scheme~\cite{Hill1952}, we take an arithmetic average of the Voigt and Reuss bounds.

\begin{equation}
    \text{Bulk modulus},~~ 
    K_{VRH} = \frac{K_{V} + K_{R}}{2}
    \label{eq:Kvrh}
\end{equation}

\begin{equation}
    \text{Shear modulus},~~ 
    G_{VRH} = \frac{G_{V} + G_{R}}{2}
    \label{eq:Gvrh}
\end{equation}

From the knowledge of $C_{ij}$, bulk ($K$) and shear ($G$) moduli, one can estimate various other material-specific parameters, as listed below 

\begin{equation}
    \text{Young's modulus},~~ 
    E = \frac{9KG}{3K + G}.
    \label{eq:E}
\end{equation}

\begin{equation}
    \text{Isotropic Poisson's ratio},~~ 
    \nu = \frac{3K - E}{6K} \equiv \frac{3K-2G}{2(3K+G)}.
    \label{eq:nu}
\end{equation}

\begin{equation}
    \text{P-wave modulus},~~ 
    M = K + \frac{4G}{3}.
    \label{eq:pwave}
\end{equation}

\begin{equation}
    \text{Lame's first parameter~\cite{mavko2020}},~~ 
    \lambda = \frac{\nu E}{(1+\nu)(1-2\nu)}.
    \label{eq:lame1}
\end{equation}

\begin{equation}
    \text{Lame's second parameter~\cite{mavko2020}},~~ 
    \mu = \frac{E}{2(1+\nu)}.
    \label{eq:lame2}
\end{equation}

\begin{equation}
    \text{Pugh's ratio~\cite{pugh1954}} = \frac{K}{G}.
    \label{eq:pughs}
\end{equation}

\begin{equation}
    \text{Kleinman's parameter~\cite{Kleinman1962, harrison2012}},~~ 
    \zeta = \frac{C_{11} + 8C_{12}}{7C_{11} - 2C_{12}}.
    \label{eq:klein}
\end{equation}

\begin{equation}
    \text{Cauchy's pressure},~~ 
    P_{C} = C_{12}-C_{44}.
    \label{eq:CP}
\end{equation}

In the following, we present a brief introduction to the moduli defined in Eqs.~\ref{eq:pwave}--\ref{eq:CP}. 
The pressure wave or P-wave modulus, also known as the longitudinal modulus, is associated with the homogeneous isotropic linear elastic materials and it describes the ratio of axial stress to the axial strain in a uniaxial strain state~\cite{mavko2020}.

Lame's first and second parameters help us to parameterize the Hooke's law in 3D for homogeneous and isotropic materials using the stress and strain tensors. The first parameter $\lambda$ provides a measure of the compressibility, whereas the second parameter $\mu$ is associated with the shear stiffness of a given material~\cite{mavko2020}. 

Pugh's ratio defines the ductility or brittleness of a given material. The critical value of Pugh's ratio is found to be 1.75. Materials with $K/G > 1.75$ are ductile, whereas those with $K/G < 1.75$ are brittle in nature~\cite{Pavlic2017, SinghPRB2018, pugh1954, SinghPRB_rapid2016}. 

Kleinman’s parameter ($\zeta$) describes the stability of a material under bending or stretching. A low/high value of $\zeta$ indicates that system is less/more resistant to bond bending. $\zeta$ = 0 implies that bond bending would be dominated, while $\zeta$ = 1 implies that bond stretching would be dominated~\cite{Kleinman1962, harrison2012}.

Cauchy’s pressure ($P_{C}$) is associated with the angular characteristic of atomic bonding in a given material. A positive value of $P_{C}$ dictates the dominance of ionic bonding, while a negative value of $P_{C}$ dictates the dominance of covalent bonding.

\subsection{Elastic Anisotropy}
Elastic anisotropy is an important property to characterize for a comprehensive understanding of the physical and mechanical properties of crystals~\cite{Hearmon1946, zener1948, ChungJAP1967, RanganathanPRL2008, Kube2016, Ledbetter2006, KubeMaarten2016}. Even the highly-symmetric cubic crystals show substantial directional dependence of the elastic properties~\cite{zener1948}. Various different methods~\cite{zener1948, ChungJAP1967, RanganathanPRL2008, Kube2016}, as listed below, have been reported in literature to quantify the elastic anisotropy from the knowledge of elastic moduli and $C_{ij}$ tensor. \\ 

Zener's anisotropy (for cubic crystals only)~\cite{zener1948}  
\begin{equation}
    A_{Z} = \frac{2C_{44}}{C_{11} - C_{12}}.
    \label{eq:AZ}
\end{equation}

Chung-Buessem empirical anisotropy index (for cubic crystals only)~\cite{ChungJAP1967}
\begin{equation}
    A_{CB} = \frac{G_{V} - G_{R}}{G_{V} + G_{R}}.
    \label{eq:Acb}
\end{equation}

Ranganathan and Ostoja-Starzewski universal anisotropy index (for any locally isotropic crystal)~\cite{RanganathanPRL2008} 
\begin{equation}
    A_{U} = \frac{K_{V}}{K_{R}} + 5\frac{G_{V}}{G_{R}} - 6.
    \label{eq:AU}
\end{equation}

Kube's log-Euclidean anisotropy index (for any crystal class)~\cite{Kube2016} 
\begin{equation}
    A^{L} = \sqrt{\left [ ln \left ( \frac{K_{V}}{K_{R}} \right ) \right ]^{2} + 5 \left [ln \left( \frac{G_{V}}{G_{R}} \right)\right ]^{2}} .
    \label{eq:AL}
\end{equation}

Zener's anisotropy~\cite{zener1948} was the first attempt at quantifying anisotropy in cubic crystals. Zener proposed a comparison of the ratio of shear resistances of the on-axis and the off-axis terms, this produced the term shown in Eq.~\ref{eq:AZ}. 
Whenever the resistances are equal ({\it i.e.}, isotropic case) equation Eq.~\ref{eq:AZ} produces unity, and any deviation from unity can be regarded as a measure of the anisotropic behavior of the crystal under investigation. The shortcomings of this equation are: (i) only the cubic elastic tensor was considered, and (ii) it is a non-unique definition, $i.e.,$ the equation could have easily been defined as its inverse instead. In literature, this was a starting point for an anisotropy measure, but more recently more generalized definitions of elastic anisotropy are used since they can be used on other crystal classes and are uniquely defined~\cite{ChungJAP1967,RanganathanPRL2008,Kube2016}. 

Chung-Buessem anisotropy~\cite{ChungJAP1967} was defined to obtain ``single-valuedness" for cubic crystals, which allows for relative comparison of cubic crystals. In  Eq.~\ref{eq:Acb}, $A_{CB} = 0$ for isotropic crystals and any positive deviation from this limiting value would indicate an anisotropic behaviour. With this definition one can say for certain whether a given cubic crystal is more anisotropic than the other. 

The evolution of the definition of elastic anisotropy eventually led to the need of its generalization for any crystal system. The universal anisotropy index  (Eq.~\ref{eq:AU}) was proposed for this purpose~\cite{RanganathanPRL2008}. This generalization takes into account the tensorial nature of the elastic tensor and as well as all the elastic moduli. Similar to the Chung-Buessem anisotropy, the universal anisotropy index~\cite{RanganathanPRL2008} defines the limiting value of isotropic crystals to be zero, and any  positive deviations from zero would indicate the elastic anisotropy. 
 One must take care in this definition as it only provides relative comparison between two crystals; it does not allow for definitive comparison such as: one crystal is twice as anisotropic as another.

The Kube's log-Euclidean anisotropy~\cite{Kube2016} is the most general definition of anisotropy at present, as it was devised to make definitive comparisons between any two crystals. This is done by defining a distance measure between the Ruess and Voigt estimation of elastic constants. Log-Euclidean anisotropy is very similar in behavior to the previous two anisotropies. Here, isotropy is determined by $A^{L} = 0$ and any positive value denote a measure of the elastic anisotropy. Now, thanks to this definition, one can definitively say that one crystal is twice as anisotropic as another one.




\subsection{Elastic Wave Velocities}
The longitudinal ($v_{l}$), transverse ($v_{t}$), and average ($v_{m}$) elastic wave velocities can be estimated from the knowledge of the bulk and shear moduli, and the density of material ($\rho$)~\cite{ANDERSON1962, screiber1973elastic}.

\begin{equation}
v_{l} = \sqrt{\frac{3K + 4G}{3\rho}},
\label{eq:vl}
\end{equation}

\begin{equation}
v_{t} = \sqrt{\frac{G}{\rho}}, \text{~~and}
\label{eq:vt}
\end{equation}

\begin{equation}
v_{m} =  \bigg[\frac{1}{3}\bigg(\frac{2}{v_{t}^3} + \frac{1}{v_{l}^3}\bigg)\bigg]^{-1/3}. 
\label{eq:vm}
\end{equation}

The above equations also imply that one can determine the elastic moduli and elastic stiffness constants by measuring the elastic wave velocities using ultrasonic waves~\cite{Gopinathan_1974, Lichnowski_1976, Varkey1978, Maciej2009}.
The {\sc MechElastic} package uses the VRH average of $K$ and $G$ values for calculation of $v_{l}$ and $v_{t}$.  

\subsection{Debye Temperature}
Debye temperature ($\Theta_{D}$) is an important thermodynamic parameter that can be estimated from the knowledge of the average elastic wave velocity $v_{m}$ and the density of material $\rho$. We note that at low-temperatures  $\Theta_{D}$ estimated from the elastic constants is approximately same as the $\Theta_{D}$ obtained from the specific heat measurements. This is due to the fact that at low-temperatures the acoustic phonons are the only vibrational excitations contributing to the specific heat of material. We calculate $\Theta_D$ using the following expression~\cite{ANDERSON1962}:
\begin{equation}
\Theta_D = {\frac{h}{k_B}\bigg[\frac{3q}{4{\pi}}\frac{N\rho}{M} \bigg]^{1/3}}v_m, 
\end{equation}
where $h$ is the Planck's constant, $k_B$ is the Boltzmann's constant, $q$ is the total number of atoms in cell, $N$ is the Avogadro's number, $\rho$ is the density, and $M$ is the molecular weight.

\subsection{Melting Temperature}
The melting temperature ($T_{melt}$) is estimated using the below empirical relation~\cite{FINE1984951, Johnston1996}, which is a very crude approximation and its validity must be checked carefully for the material under investigation.
\begin{equation}
T_{melt} = 607 + 9.3~K_{VRH}\, \pm\, 555.
\label{eq:melt}
\end{equation}

\begin{figure}[htb!]
    \centering
    \includegraphics[width=0.7\linewidth]{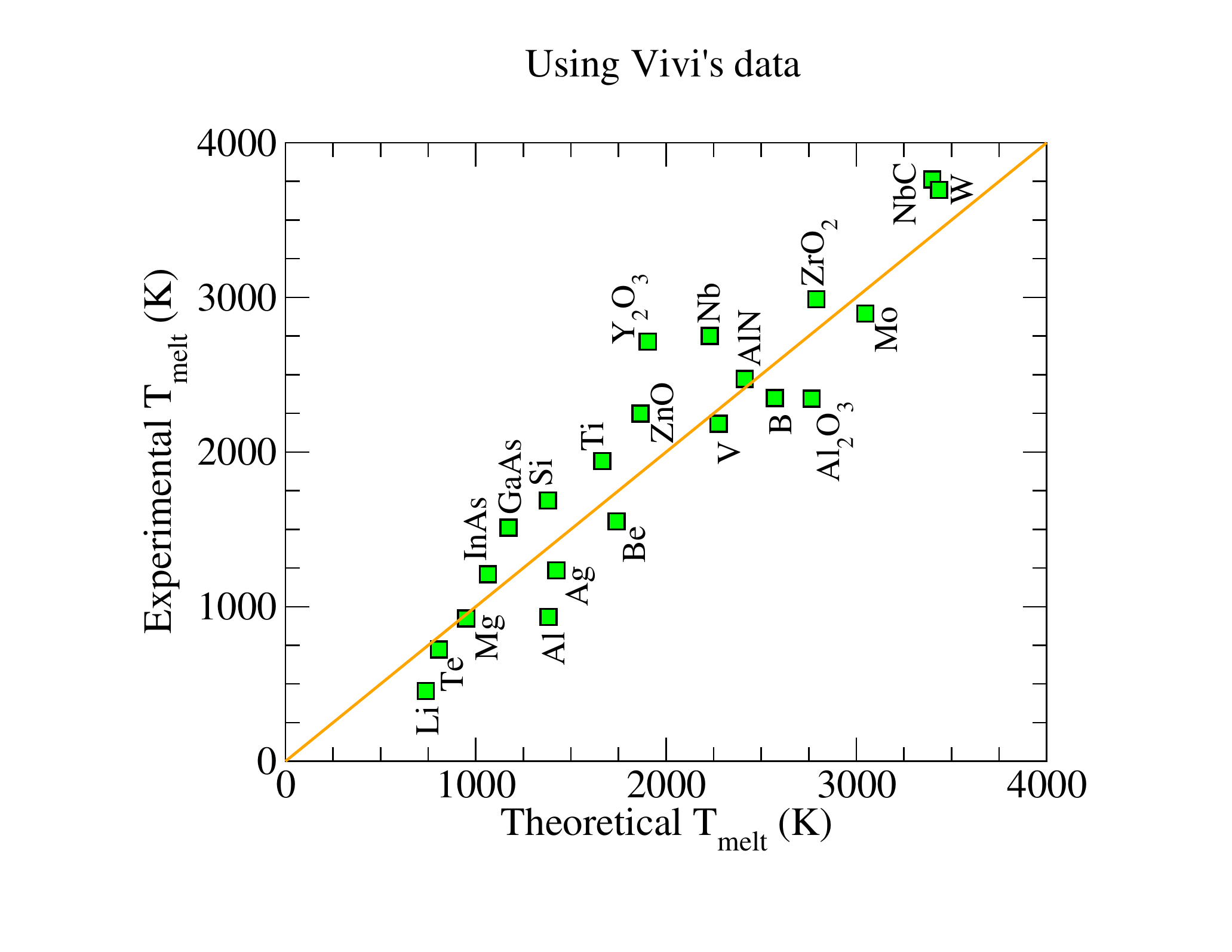}
    \caption{A comparison of the theoretically estimated (using Eq.~\ref{eq:melt}) and experimentally reported melting temperature ($T_{melt}$) data for some elemental and binary compounds. All the structures and their elastic-tensor data were obtained from the Materials Project database~\cite{Jain2013}, and the experimental data were obtained from the website: \href{https://www.americanelements.com/}{https://www.americanelements.com/}. The mp ID of the structures is as follows: Li (mp-135), Te (mp-19), Mg (mp-153), InAs (mp-20305), GaAs (mp-2534), Si (mp-149), Al (mp-134), Ag (mp-124), Ti (mp-46), Be (mp-87), ZnO (mp-2133), Y$_2$O$_3$ (mp-2652), Nb (mp-75), V (mp-146), AlN (mp-661), B (mp-160), Al$_2$O$_3$ (mp-1143), ZrO$_2$ (mp-1565), Mo (mp-129), NbC (mp-910), and W (mp-91).  }
    \label{fig:tmelt_compare}
\end{figure}

In order to test the validity of Eq.~\ref{eq:melt}, we estimated $T_{melt}$ for several materials using the elastic-tensor data obtained from the Materials Project database~\cite{Jain2013}, and compared it with the experimental values, as shown in Figure~\ref{fig:tmelt_compare}. As one can notice, there is a definite correlation between the theoretical and experimental values, however, the uncertainty in the theoretical estimates of $T_{melt}$ is substantial. Therefore, one should be careful in using Eq.~\ref{eq:melt} for theoretical estimation of $T_{melt}$. Also, note that the units of $K_{VRH}$ and $T_{melt}$ in the above empirical relation are in GPa and K, respectively.

\section{Hardness} 
\label{sec:hardness}
DFT has become a powerful tool for calculating the elastic tensor of materials, allowing the successful estimation of mechanical properties such as bulk ($K$), shear ($G$), Young's moduli ($E$), and Poisson's ratio ($\nu$). Nevertheless, first-principles calculations are unable to predict hardness directly. Since hardness is a fundamental property that is essential to fully describe the mechanical behaviour of a solid, various semi-empirical relations have been proposed to estimate hardness using the commonly known elastic moduli. The {\sc MechElastic} package has  implementation of six different semi-empirical relations, as given below with their respective sources, to calculate Vickers hardness (GPa units) from the macroscopic elastic parameters $K$, $G$, $E$, and $\nu$. These semi-empirical relations  are nicely summarized by Ivanovskii in Ref.~\cite{2013-ivanovskii}. 

\begin{equation}
H_{1a} = 0.1475\,G ~~~~
\text{Ref.~\cite{H1-jiang}}
\end{equation} 

\begin{equation}
H_{1b} = 0.0607\,E ~~~~
\text{Ref.~\cite{H1-jiang}}
\end{equation}

\begin{equation}
H_2 = 0.1769\,G -2.899 ~~~~
\text{Ref.~\cite{H2-teter}}
\end{equation}

\begin{equation}
H_3 = 0.0635\,E ~~~~
\text{Ref.~\cite{H3-jiang}}
\end{equation}

\begin{equation}
H_4 = \frac{(1-2\,\nu)\,K}{6\,(1+\nu)} ~~~~
\text{Ref.~\cite{H4-miao}}
\end{equation}

\begin{equation}
H_5 = 2(k^2\,G)^{0.585} - 3 ~~~~
\text{Ref.~\cite{H5-chen}}
\end{equation}
The $k$ parameter in $H_5$ corresponds to the inverse of the Pugh's ratio ($G/K$).

\subsection{Hardness recommendation model}
\label{sec:recomhardness}

In order to figure out the aptness of the above-mentioned methods in the prediction of hardness for different type of materials, we calculated the Vickers hardness for more than fifty different materials, and compared the obtained theoretical values with the available experimental data. 
The elastic tensor for each material was extracted from the Materials Project's database~\cite{Jain2013}, and the mechanical properties ($K$, $G$, $E$, and $\nu$) were estimated using the {\sc MechElastic} package. 
The purpose was to create a guide for users to select the most appropriate method for the hardness calculation based on the specific characteristics of each material. 
Information of crystal class, energy bandgap ($E_{g}$), and  density and the presence of specific atoms in the crystal (like C, B, N, O and transition metals) was analyzed, and correlated to minimize the absolute error in the hardness calculation (a more comprehensive study would be published elsewhere). 
The best model was found to be the one that correlates the crystal class and the energy bandgap. For this study the materials were classified as insulators ($E_{g} >$ 2 eV), semiconductors ($0 < E_{g} <$ 2 eV) and metals ($E_{g}$ = 0 eV). Table~\ref{hardness_model} presents a guide to select the best method to compute hardness for different type of materials.

\begin{table}[h!]
\centering
\caption{A guide to select the best hardness calculation method as a function of the crystal class and bandgap}
\label{hardness_model}
\begin{tabular}{lccccc}
\hline
 & Cubic & Hexagonal & Orthorhombic & Rhombohedral & General \\
\hline
Insulator & H$_2$ & H$_{1b}$ & H$_2$ & H$_2$ & H$_2$ \\
Semiconductor & H$_5$ & H$_{1b}$,H$_3$ &  & H$_2$ & H$_5$ \\
Metal & H$_{1a}$ & H$_4$ & H$_4$ & H$_4$ & H$_4$ \\
\hline
\end{tabular}
\end{table}

\begin{figure}[htb!]
    \centering
    \includegraphics[width=0.7\linewidth]{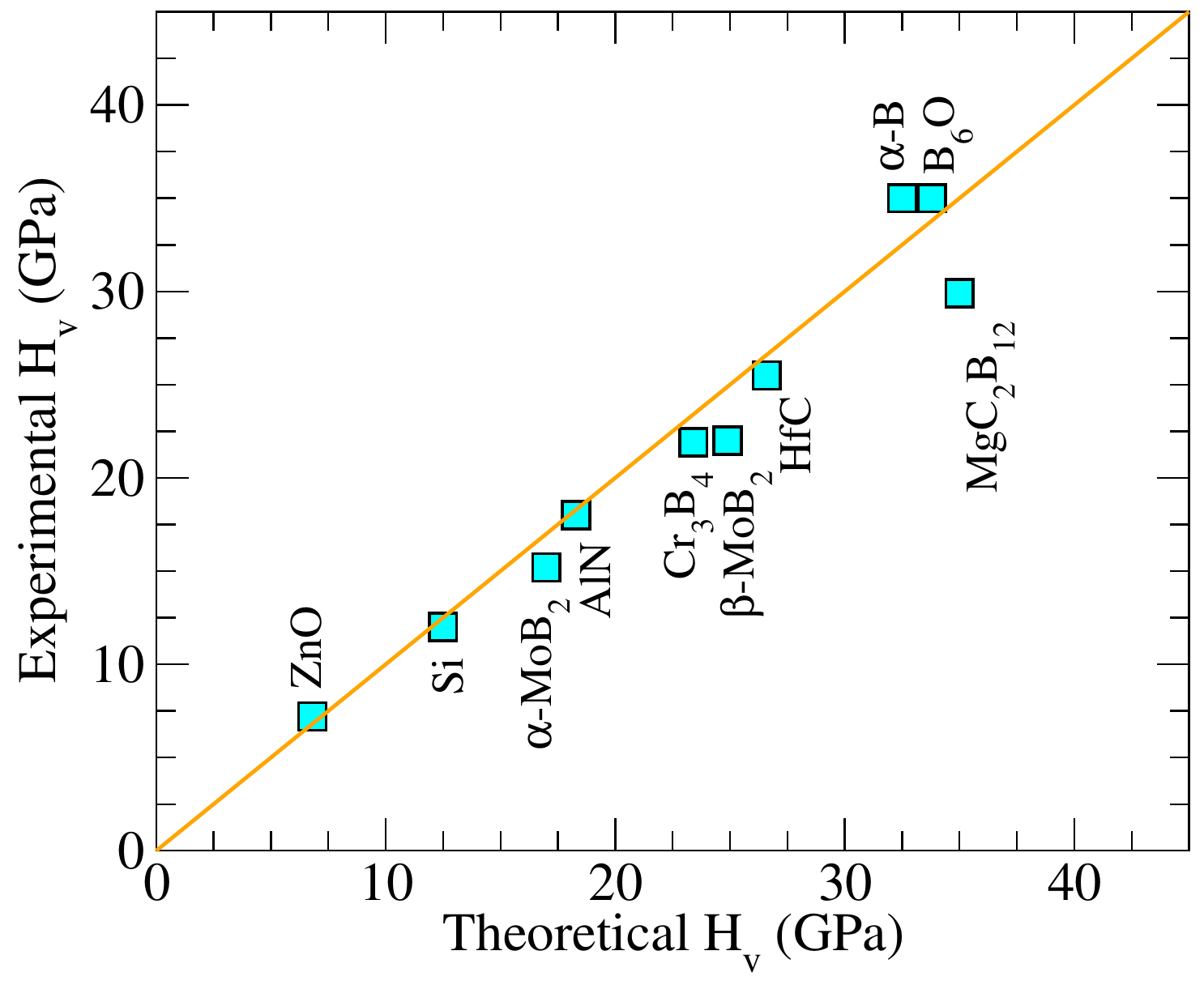}
    \caption{A comparison of the available experimental Vickers hardness (H$_v$) data with the theoretical values obtained using the recommended model in Section~\ref{sec:recomhardness} for several materials listed in Table~\ref{examples}. }  
    \label{fig:hardness_compare}
\end{figure}

Table~\ref{examples} presents the mechanical properties of several materials calculated using the {\sc MechElastic} package. In this investigation we used the elastic tensor provided in the Materials Project database~\cite{Jain2013} for each material and compared the theoretically estimated hardness values with the  experimental data reported in the literature. The hardness values were calculated using the above-mentioned recommendation model (see Table~\ref{hardness_model}).  Figure~\ref{fig:hardness_compare} illustrates a comparison between the theoretical and experimental values of the Vickers hardness for materials listed in Table~\ref{examples}. Note that the data for diamond was removed in  Fig.~\ref{fig:hardness_compare} for clarity. 

\begin{table}[h!]
\centering
\caption{Bulk modulus (GPa), Young’s modulus (GPa), shear modulus (GPa), and Poisson’s ratio calculated with the {\sc MechElastic} package using the elastic tensor provided in the Materials Project database for several materials. The bandgap (eV) extracted from the Materials Project database was used to select the recommendation model to compute the Vickers hardness (H$_v$ in GPa). The experimental values are given in parenthesis.}
\label{examples}
\resizebox{\textwidth}{!}{%
\begin{tabular}{lllllllllll}
\hline
 Material & mp-ID & Crystal class & Bandgap & Bulk & Shear & Young & Poisson & $H_v$ & $H_{method}$ & Ref. \\
\hline
Diamond & mp-66 & Cubic & 4.34 & 435 (446) & 521 (535) & 1117 (1143) & 0.07 (0.07) & 89.2   (96.0) & H$_2$ & \cite{H5-chen,H8-young} \\
\ce{Si} & mp-149 & Cubic & 0.85 & 83 (94) & 61 (63) & 148 (166) & 0.20 & 12.5   (12.0) & H$_5$ & \cite{H1-jiang,H5-chen} \\
\ce{HfC} & mp-21075 & Cubic & 0.00 & 236 (243) & 180 (180) & 431 (433) & 0.20 & 26.6 (25.5) & H$_{1a}$ & \cite{H1-jiang} \\
\ce{AlN} & mp-661 & Hexagonal & 4.05 & 195 (202) & 122 (115) & 302 (290) & 0.24 & 18.3 (18.0) & H$_{1b}$ & \cite{H1-jiang} \\
\ce{ZnO} & mp-2133 & Hexagonal & 0.73 & 130 (148) & 41 (41) & 112 (110) & 0.36 & 6.8 (7.2) & H$_{1b}$ & \cite{H1-jiang} \\
$\alpha$-\ce{MoB2} & mp-960 & Hexagonal & 0.00 & 299 (304) & 153 (180) & 392 (451) & 0.28 (0.25) & 17.0 (15.2) & H$_4$ & \cite{H7-tao} \\
\ce{MgC2B12} & mp-568803 & Orthorhombic & 2.29 & 229 (228) & 214 (217) & 489 (494) & 0.14 (0.14) & 35.0 (29.9) & H$_2$ & \cite{H3-jiang} \\
\ce{Cr3B4} & mp-889 & Orthorhombic & 0.00 & 299 (299) & 210 (215) & 511 (520) & 0.21 (0.21) & 23.4 (21.9) & H$_4$ & \cite{H4-miao} \\
\ce{B6O} & mp-1346 & Rhombohedral & 2.21 & 227 (228) & 208 (204) & 477 (471) & 0.15 & 33.8 (35.0) & H$_2$ & \cite{H1-jiang} \\
$\alpha$-Boron & mp-160 & Rhombohedral & 1.52 & 211 (224) & 200 (205) & 456 (470) & 0.14   (0.12) & 32.5 (35.0) & H$_2$ & \cite{H1-jiang,H3-jiang} \\
$\beta$-\ce{MoB2} & mp-2331 & Rhombohedral & 0.00 & 295 (298) & 225 (215) & 537 (520) & 0.20 (0.21) & 24.9   (22.0) & H$_4$ & \cite{H7-tao} \\
\hline
\end{tabular}
}
\end{table}

\section{Elastic Properties of 2D Materials}
\label{sec:2d}
In order to evaluate the elastic and mechanical properties of 2D materials, one requires to convert the 3D $C_{ij}^{3D}$ tensor obtained for a bulk unit cell ($i.e.$, a 2D layer plus vacuum added to avoid the periodic interactions) from a DFT calculation into a 2D $C_{ij}^{2D}$ tensor. This can be done by multiplying the $C_{ij}^{3D}$ with the vacuum thickness $t$, $i.e.,$ $C_{ij}^{2D}$ = $tC_{ij}^{3D}$. $t$ can be  provided by the user, else it would be automatically estimated by the {\sc MechElastic} package when `\texttt{-d\,=\,2D}' option is specified for the 2D case. 
By default, {\sc MechElastic} assumes that vacuum is added along the $c$-axis. 
Moreover, {\sc MechElastic} automatically converts the bulk units of $C_{ij}^{3D}$ from GPa (or kBar) to N/m for 2D case. 

Once $C_{ij}^{2D}$ ($\equiv c_{ij}$) is determined, one can estimate various 2D elastic moduli using the expressions given below~\cite{AELAS2017, AndrewPRB2012, Peng2012, PENG2013, SinghPRB2017_2D, AKINWANDE2017, SinghPRB2018_gMoS2}. Here, we use the lowercase letter $c_{ij}$ to denote the 2D elastic constants. Note that only four $c_{ij}$ ($i.e., c_{11}, c_{22}, c_{12},$ and $c_{66}$) are relevant for 2D square, rectangular, or hexagonal lattices, where symmetry of the square and hexagonal lattices enforces $c_{11} = c_{22}$, and $c_{66} = \frac{1}{2}(c_{11}-c_{12})$. In 2D oblique lattices having no symmetry, two additional coefficient $c_{16}$ and $c_{26}$ are allowed to remain nonzero~\cite{Ma_dziarz_2019}. \\

2D layer modulus is defined as~\cite{AndrewPRB2012}:
\begin{equation}
    \gamma = \frac{1}{4}(c_{11} + c_{22} + 2c_{12})
    \label{eq:layermod}. 
\end{equation}

2D Young’s moduli ($i.e.,$ in-plane stiffness) for strains in the
Cartesian [10] and [01] directions are~\cite{AndrewPRB2012}:   
\begin{equation}
    Y_{[10]} = \frac{c_{11}c_{22} - c_{12}^{2}}{c_{22}},~~\text{and~~} 
    Y_{[01]} = \frac{c_{11}c_{22} - c_{12}^{2}}{c_{11}}.
    \label{eq:2dyoung}
\end{equation}

2D Poisson’s ratio in the Cartesian [10] and [01] directions~\cite{AndrewPRB2012}:   
\begin{equation}
    \nu_{[10]} = \frac{c_{12}}{c_{22}},~~\text{and~~} 
    \nu_{[01]} = \frac{c_{12}}{c_{11}}.
    \label{eq:2dpoisson}
\end{equation}

2D shear modulus is defined as~\cite{AndrewPRB2012}:   
\begin{equation}
    G = c_{66}. 
    \label{eq:2dpoission}
\end{equation}

Since the above-mentioned properties are well defined only for the 2D materials, 
they require a re-scaling for quasi-2D systems having a considerable finite thickness along the 
out-of-plane direction. One can convert the 2D units of N/m into the bulk unit 
of N/m$^{2}$ by dividing the elastic moduli by the thickness of material. 

\subsection{2D Mechanical Stability Criteria and Role of Crystal Symmetries}
With the $c_{ij}$ matrix we are able to test the mechanical stability of an unstressed 2D lattice using the  conditions presented by Marcin Maździarz in Ref.~\cite{Ma_dziarz_2019}. 
Similar to the 3D case, mechanical stability in 2D is not just determined by a check of the positive definiteness of the $c_{ij}$ matrix. 

In Ref.~\cite{Ma_dziarz_2019}, Maździarz presented the mechanical stability criteria for five Bravais lattices, which are implemented in the {\sc MechElastic} package. 
The general form of $c_{ij}$ matrix for any 2D lattice can be written as
\vspace{0.4cm}

\begin{equation}
\begin{bmatrix}
 C_{11}   &   C_{12}    & C_{16}     \\
 \bullet  &   C_{22}     & C_{26}     \\
 \bullet  &  \bullet     &  C_{66}     \\
\end{bmatrix}.
\label{2d_cij}
\end{equation}
\vspace{0.4cm}

\subsubsection{Square Lattice Conditions}
\begin{equation}
\begin{split}
C_{11} - C_{12} > 0\,~~  \\
C_{11} + C_{12} > 0\,~~  \\ 
C_{66} > 0\,~~  \\ 
\end{split}
\label{eq:2d_square_stability}
\end{equation}

\subsubsection{Rectangular Class Lattice Conditions}
\begin{equation}
\begin{split}
\frac{1}{2}(C_{11} + C_{22} + \sqrt{4C_{12}^2 - (C_{11} - C_{22})^2}) > 0\,~~  \\
\frac{1}{2}(C_{11} + C_{22} - \sqrt{4C_{12}^2 - (C_{11} - C_{22})^2}) > 0\,~~  \\ 
C_{66} > 0\,~~  \\ 
\end{split}
\label{eq:2d_rectangular_stability}
\end{equation}

\subsubsection{Hexagonal Lattice Conditions}
\begin{equation}
\begin{split}
C_{11} - C_{12} > 0\,~~  \\
C_{11} + C_{12} > 0\,~~  \\ 
\end{split}
\label{eq:2d_hexagonal_stability}
\end{equation}

\subsubsection{Oblique Lattice Conditions}
\begin{equation}
\begin{split}
C_{11} > 0\,~~  \\
C_{11}C_{22} > C_{12}^2\,~~  \\ 
det(C_{ij}) > 0 \,~~  \\ 
\end{split}
\label{eq:2d_oblique_stability}
\end{equation}

\section{ELATE Implementation}
\label{sec:elate}
ELATE is a popular online elastic tensor analysis tool developed by Gaillac and Coudert~\cite{elate,elate2016}, which takes an elastic tensor as input, and subsequently calculates the spatial variation of the Young's and shear moduli, the linear compressibility, and the Poisson's ratio. With the addition of the mass density, it can also calculate the variations in the compressional and shear sound speed, the ratio of the speeds, and also an estimate of the Debye speed. Along with the variations of the properties, the package calculates the spatial average of all the aforementioned properties. 
In ELATE, the mechanical stability is tested by finding the eigenvalues of the elastic tensor and checking if they all are positive. 
A concise summary of the spatial variations of the Young's and shear moduli, linear compressibility, and Poisson's ratio is given below. 
We refer the reader to Ref.~\cite{elate2016} for a more detailed information. In this summary, the minimum and the maximum values, and a measurement of the anisotropy for each property are given along with the direction for which they occur. The spatial variations of the properties are then visualized with 2D contour plots and 3D plots in the elastic property space. The color scheme of these plots are green, red, and blue, which corresponds to the minimal positive value, the minimal negative value, and the maximum possible positive values these properties can take. The 2D contours are taken along the primary axes, whereas the 3D plots belong to the full surface.

In {\sc MechElastic}, ELATE is converted to an accessible offline tool in Python. An object called ELATE is constructed, which can be used to access the elastic data as the properties of the object. With this object the method \texttt{print\_properties()} can be used to produce the same concise analysis as the original results obtained by {\sc MechElastic}. 
The data for the 2D and 3D plots are generated in the same way as done in the original ELATE software, but are returned as tuples for direct user access. This gives the user freedom to manipulate the data to create their own graphs and figures. These plots are produced by functions \texttt{plot\_2D()} and \texttt{plot\_3D()} of the object, and are plotted using the packages \texttt{matplotlib}~\cite{matplotlib} and \texttt{pyvista}~\cite{pyvista}, respectively. The color scheme of the plots have the same meaning as in the original ELATE software~\cite{elate, elate2016}.

\section{Equation of State Analysis}
\label{sec:eos}

An Equation of State (EOS) is a thermodynamic relation between the state variables such as temperature (T), volume (V), pressure (P), internal energy, and/or specific heat, etc. It helps us to describe the state of a system under given physical conditions.
For instance, it could be used to study a certain property of a material such as its density, under varying pressure or temperature or any other physical variable. The EOS's are used in a wide range of fields ranging from Geo-Physics to Materials Science~\cite{anderson1996equations}. Other than predicting thermodynamical properties, EOS can be used to gain insight into the nature of solid-state and molecular theories \cite{vinet1987compressibility}. 

The EOS models considered here are for the isothermal processes, where T is kept constant while P and V could be varying. Experimentally, these values can be obtained through x-ray diffraction (XRD), Diamond Anvil Cells (DAC), or shock experiments. 
However, the pressure range obtainable by such experimental techniques is limited and therefore, it calls for theoretical methods to extend the range for environments in extreme conditions such as planetary interiors.   

{\sc MechElastic}'s EOS class contains methods to plot `Energy vs. Volume' and `Pressure vs. Volume' curves from energy, pressure and volume data obtained from either calculations or experiments using several fitting models including Vinet, Birch, Murnaghan, and Birch-Murnaghan~\cite{birch1938effect, birch1947finite, murnaghan1951finite, birch1952elasticity, birch1977isotherms, birch1978finite, vinet1987compressibility, vinet1989universal, stacey2001finite, Loose2013}. 
As calculations are sensitive to the type of the EOS model, outcomes from multiple models should be carefully examined~\cite{hebbache_ab_2004}. {\sc MechElastic} employs a central-difference scheme to calculate the pressure from the `Energy vs. Volume' data, and similarly an integration scheme to calculate the energy from `Pressure vs. Volume' data. This is especially helpful for studying the phase diagrams of materials to investigate the existence of phase boundaries. 

To the best of our knowledge, currently there is no universal EOS model that is applicable to all types of solids and accurate over the whole range of pressure, especially when a solid undergoes several structural phase transitions within the region of interest~\cite{gordon_condensed-matter_2016}. 
Therefore, we have included several EOS models in the {\sc MechElastic} package whose accuracy can be evaluate by the mean-squared error (MSE) of the fitting. These models are briefly described below.

\begin{figure}[h!]
  \centering
  \begin{subfigure}[b]{0.8\linewidth}
    \includegraphics[width=\linewidth]{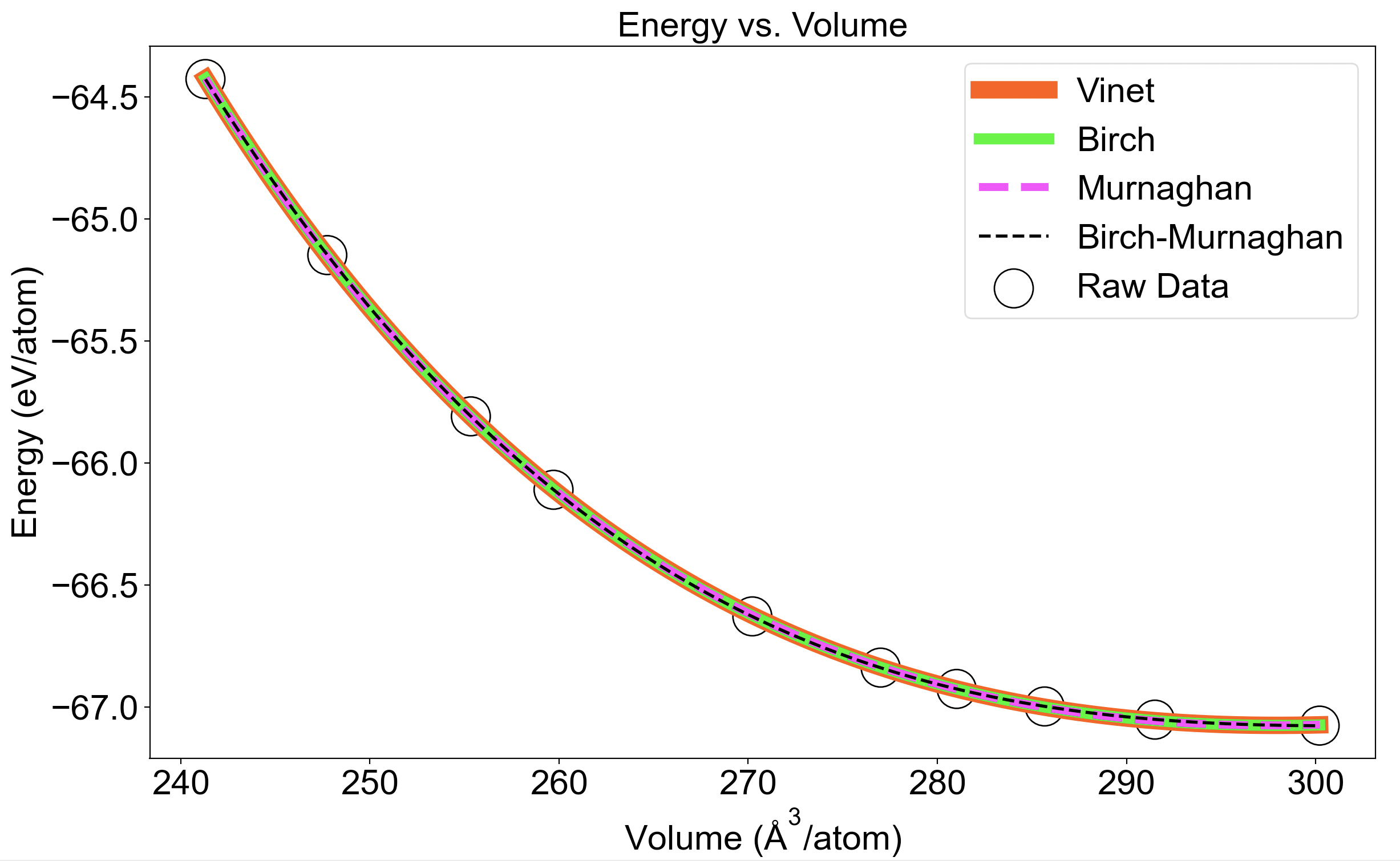}
     \caption{}
  \end{subfigure}\qquad
  \begin{subfigure}[b]{0.8\linewidth}
    \includegraphics[width=\linewidth]{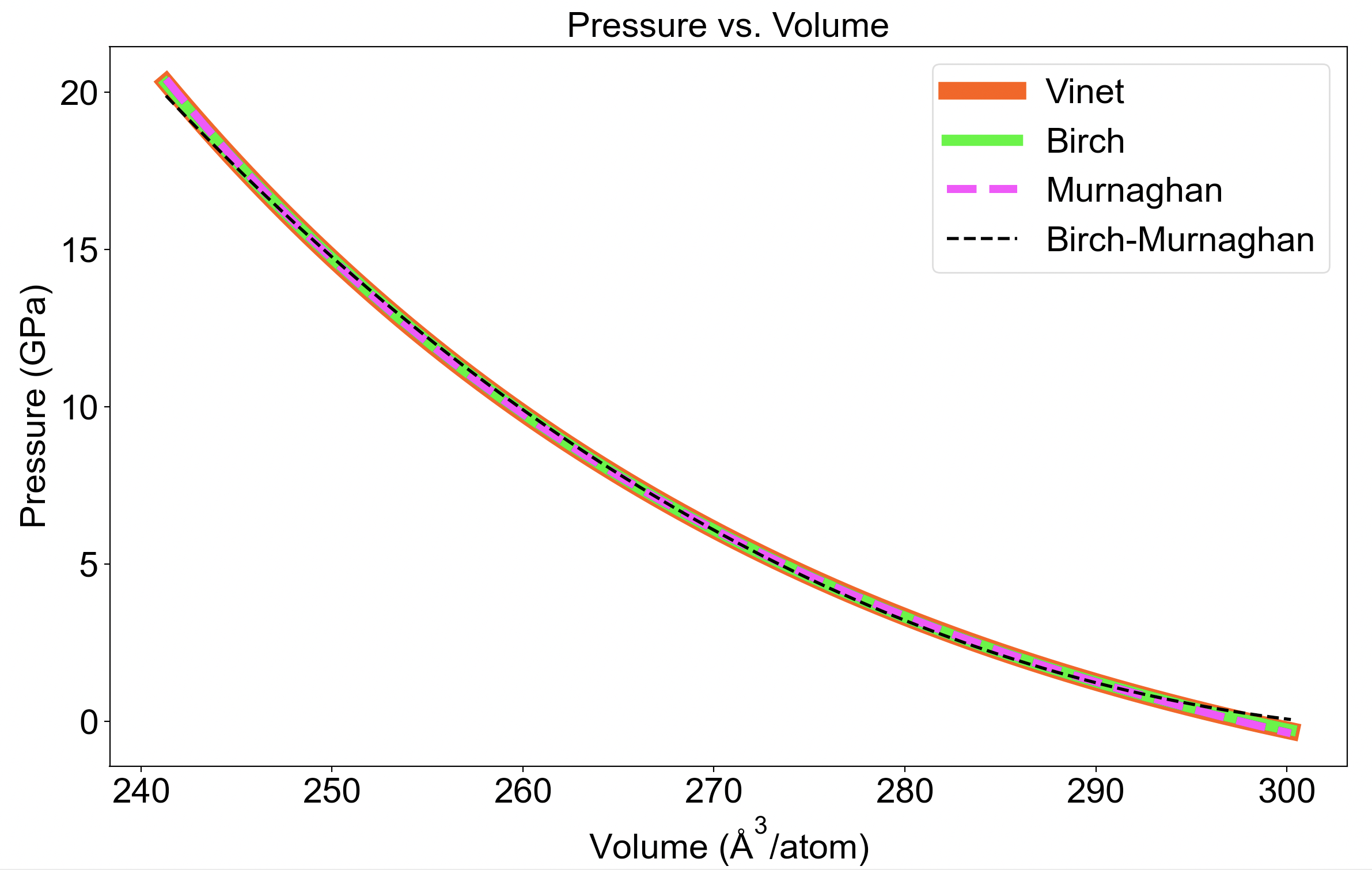}
    \caption{}
  \end{subfigure}\qquad

  \caption{The fitted equation of state (EOS) plots for (a) Energy vs. Volume and (b) Pressure vs. Volume data obtained by differentiating the energy with a central difference scheme using different EOS models. Here, the material under investigation is T$_{d}$-MoTe$_2$ for which `Energy vs. Volume' data was computed using the VASP code (numerical details are given in Ref.~\cite{SinghPRL2020}). 
  }
  \label{fig:eos_plot}
\end{figure}

\subsection{Vinet EOS}
The Vinet EOS model is based on the empirical interatomic potentials and is formulated as follows~\cite{vinet1987compressibility, vinet1989universal}: 
\begin{equation}
E(\eta)= E_{o}+\frac{2 B_{o} V_{o}}{\left(B_{o}^{\prime}-1\right)^{2}} \\
 \times\left(2-\left(5+3 B_{o}^{\prime}(\eta-1)-3 \eta\right) e^{-3\left(B_{o}^{\prime}-1\right)(\eta-1) / 2}\right)
\end{equation}

\noindent
where, $\eta=\left(V / V_{o}\right)^{1 / 3}$. $V_0$, $E_0$, $B_0$, and  $B_0^{\prime}$ denote the equilibrium volume, total energy, bulk modulus and its pressure derivative at zero pressure, respectively. 

Vinet model performs better at high pressure and high compressibility (upto 40 \%) since it includes non-linear pressure contributions as opposed to the Birch and Murnaghan EOS models~\cite{vinet1987compressibility}. However, it is not suitable for solids with significant structural flexibility, such as bond-bending in materials such as feldspars. 

\subsection{Birch EOS}

The Birch EOS model is an empirical model as written below~ \cite{michael_j_mehl_first-principles_1994}:

\begin{equation}
    \begin{aligned}
E(V)=& E_{0}+\frac{9}{8} B_{0} V_{0}\left[\left(V_{0} / V\right)^{2 / 3}-1\right]^{2}+\frac{9}{16} B_{0} V_{0}\left(B_{0}^{\prime}-4\right) \times \\
&\left[\left(V_{0} / V\right)^{2 / 3}-1\right]^{3}+\sum_{n=4}^{N} \gamma_{n}\left[\left(V_{0} / V\right)^{2 / 3}-1\right]^{n}
\end{aligned}
\end{equation}

where, E$_0$, V$_0$, B$_0$ and B$_0^{\prime}$ are the equilibrium energy, volume, bulk modulus and pressure derivative of the bulk modulus, respectively. $n$ denotes the order of the fit. 

For experimental high pressure data, Birch model fares better than the  Murnaghan model, but underperforms the Vinet model \cite{vinet_compressibility_1987,birch_elasticity_1952}.

\subsection{Murnaghan EOS}

The Murnaghan EOS model is given as below~\cite{fu_first-principles_1983}:
\begin{equation}
    E_{T}(V)=E_{T}\left(V_{0}\right)+\frac{B_{0} V}{B_{0}^{\prime}}\left[\frac{\left(V_{0} / V\right)^{B_{0}^{\prime}}}{B_{0}^{\prime}-1}+1\right]-\frac{V_{0} B_{0}}{B_{0}^{\prime}-1}
\end{equation}

where, 
V is the volume, and B$_0$ and B$_{0}^{\prime}$ denote the bulk modulus and its pressure derivative at the equilibrium volume V$_0$.

While results from the Murnaghan model agree well with the data obtained at low pressures and low compressions (upto 10 \%), they deviate from the high pressure ones~\cite{vinet1987compressibility}. The Murnaghan EOS is based on the empirical data and it does not include non-linear pressure contributions. When deriving the Murnaghan EOS, it is assumed that the bulk modulus is a linear function of pressure, $i.e.$, $K=K_{0}+P\,K_{0}^{\prime}$~\cite{fd_murnaghan_compressibility_1944}.

\subsection{Birch-Murnaghan EOS}

This model is based on the pressure expansion of the bulk modulus and finite strain theory, and consequently, it is valid only at moderate compression~\cite{hebbache_ab_2004}. This model assumes Eulerian strains under hydrostatic compression.
This EOS is continuously differentiable and higher-order terms of the Taylor expansion are negligible. The Birch-Murnaghan EOS model is the most used EOS model in Earth physics. The model for the second order is shown below.

\begin{equation}
    E(\eta)=E_{o}+\frac{9 B_{o} V_{o}}{16}\left(\eta^{2}-1\right)^{2}\left(6+B_{o}^{\prime}\left(\eta^{2}-1\right)-4 \eta^{2}\right)
\end{equation}

where, $\eta=\left(V / V_{o}\right)^{1 / 3}$ and $V_0$ , $B_0$ , $B_0^{\prime}$ and $E_0$ the total energy at zero pressure are the fitting parameters. $B_0$ , $B_0^{\prime}$ are the bulk modulus at zero pressure and its pressure derivative, respectively. 

An improvement of the Birch-Murnaghan model from other models is the inclusion of the third-order strain components~\cite{birch_finite_1947}.

\bigskip
Initially, a text file with two columns containing volume and energy or volume and pressure, respectively, should be passed into {\sc MechElastic}. With this, an initial second order parabolic \texttt{numpy}~\cite{numpy} polyfit is performed to obtain the initial fitting parameters, $E_0$ or $P_0$, $B_0$, $B_p$ and $V_0$. Afterwards, a more accurate fitting for the energy or pressure is performed using the least square method in \texttt{scipy} \cite{2020SciPy} against each EOS model. A central-difference scheme is used to obtain the pressure from this fitted energy. If pressure is to be obtained from energy, an integrating scheme is used instead. Finally, a plot of `Energy vs. Volume' and `Pressure vs. Volume' is obtained, as shown in Fig.~\ref{fig:eos_plot}, comparing the values obtained from the different EOS models with the originally provided raw data. With these plots, an analysis of the phase boundaries could be performed. A range of initial and final values for the volume can be set with the \texttt{vlim} parameter and if not given, the minimum and maximum would be calculated from the provided dataset. A desired model can be provided with the \texttt{model} parameter and if not specified, the calculation would be performed against all the available models.

\begin{figure}[htb!]
    \centering
    \includegraphics[width=1.0\linewidth]{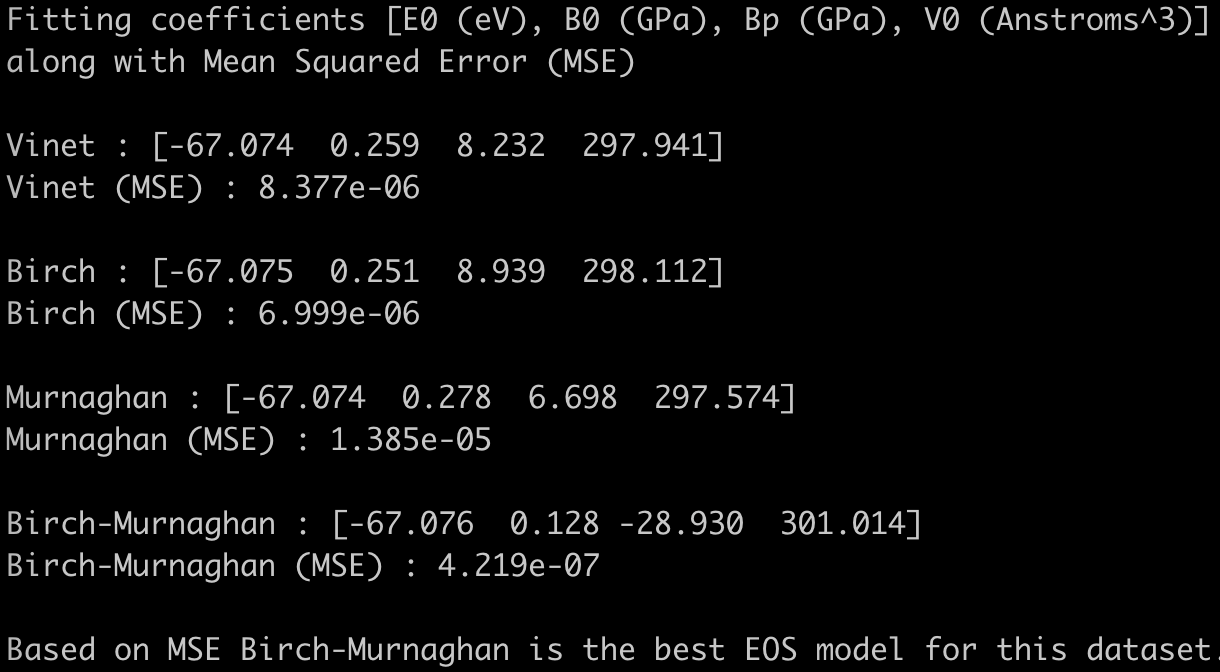}
    \caption{The output of an EOS analysis with {\sc MechElastic}. {\sc MechElastic} also prints the best EOS model to use for the provided dataset. The EOS fitting was performed for the T$_d$-MoTe$_2$ dataset, as shown in Figure~\ref{fig:eos_plot}. }
    \label{fig:eosfitoutput}
\end{figure}

In addition to the fitting, {\sc MechElastic} also performs a regression analysis for fitting using the Mean-Squared Error (MSE) of the residuals. This allows users to decide the best EOS model to use for their dataset. A sample output of the EOS analysis is shown in Fig.~\ref{fig:eosfitoutput}, which displays the fitting coefficients, E$_0$, B$_0$, B$_p$, and V$_0$ along with the Mean-Squared Error (MSE) of the residuals for each EOS model. 

\bigskip
\noindent
\textbf{Usage:}
\begin{lstlisting}[language=Python]
from mechelastic import EOS
eos_object = EOS()
eos_object.plot_eos(``energyvsvolume.dat'', eostype=``energy'', natoms=1, au=False)
\end{lstlisting}

\texttt{energyvsvolume.dat} is a text file with volume as the first column and energy as the second. \texttt{eostype} is set to \texttt{energy} for this case. If pressure and volume data is provided, \texttt{eostype} is set to \texttt{pressure} instead. \texttt{natoms} is the number of atoms in the structure and is used to output the quantities per atom. Setting \texttt{au=True} sets the units in Ha while it is in eV when False. 

\subsection{Enthalpy Curves}
Enthalpy ($H$) is defined as the sum of a system's internal energy ($U$) and the product of its pressure ($P$) and volume ($V$), $i.e.$, 
\begin{equation}
     H = U + PV
    \label{eq:enthalpy_eq}. 
\end{equation}

It is a measure of a system's capacity to do non-mechanical work and the ability to release heat.
{\sc{MechElastic}}'s EOS analysing tool is equipped with a function to calculate the `Enthalpy vs. Pressure' curves provided the energy. Pressure is calculated from the `Energy vs. Volume' data using a central-difference scheme, similar to the one mentioned in the previous section.
{\sc{MechElastic}} first automatically detects the suitable EOS model for the fitting for each phase, and then it calculates the pressure for corresponding energy values. 
The \texttt{plot\_enthalpy\_curves} method implemented within {\sc{MechElastic}} plots the `Enthalpy vs. Pressure' curves for multiple datasets of different phases and determines the phase-transition pressure by identifying the intersection points among the curves. 
The volume ranges are to be set individually for each phase, using the \texttt{vlim\_list} option. This is a list that contains elements of the minimum and maximum value of volume for each phase. 
Care must be given while selecting the volume ranges, as different models are sensitive to different volume ranges for each phase, and it may produce erroneous results if \texttt{vlim\_list} exceeds the valid volume range. If not provided {\sc{MechElastic}} will determine the best volume ranges based on the input data.

\begin{figure}[htb!]
    \begin{center}
  \includegraphics[width=0.8\linewidth]{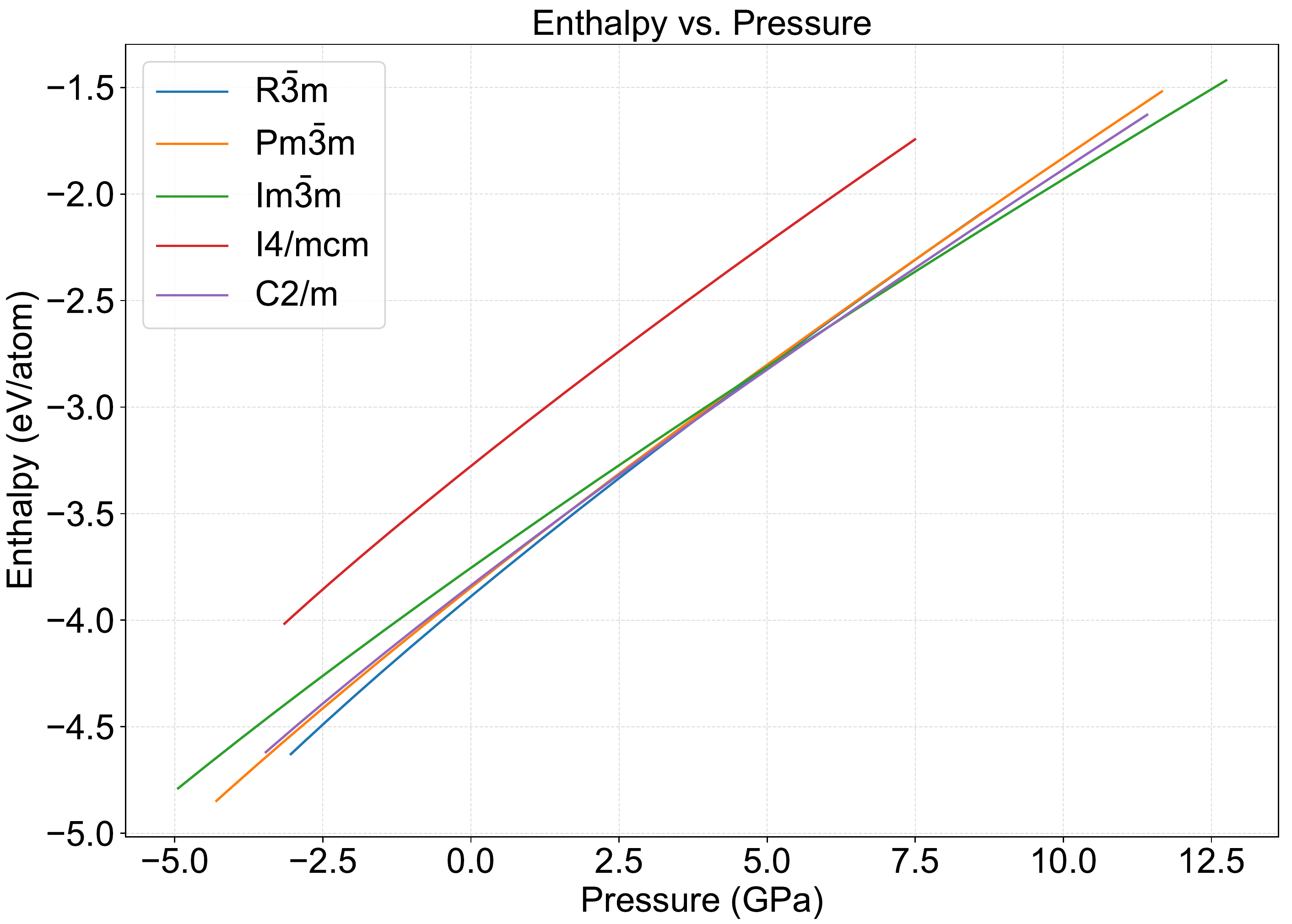}
  \caption{Enthalpy variation with pressure for five different phases of Bi as calculated using the \texttt{plot\_enthalpy\_curves} function.}  
 \label{fig:enthalpy}
     \end{center}
\end{figure}

\begin{figure}[htb!]
    \begin{center}
  \includegraphics[width=1.0\linewidth]{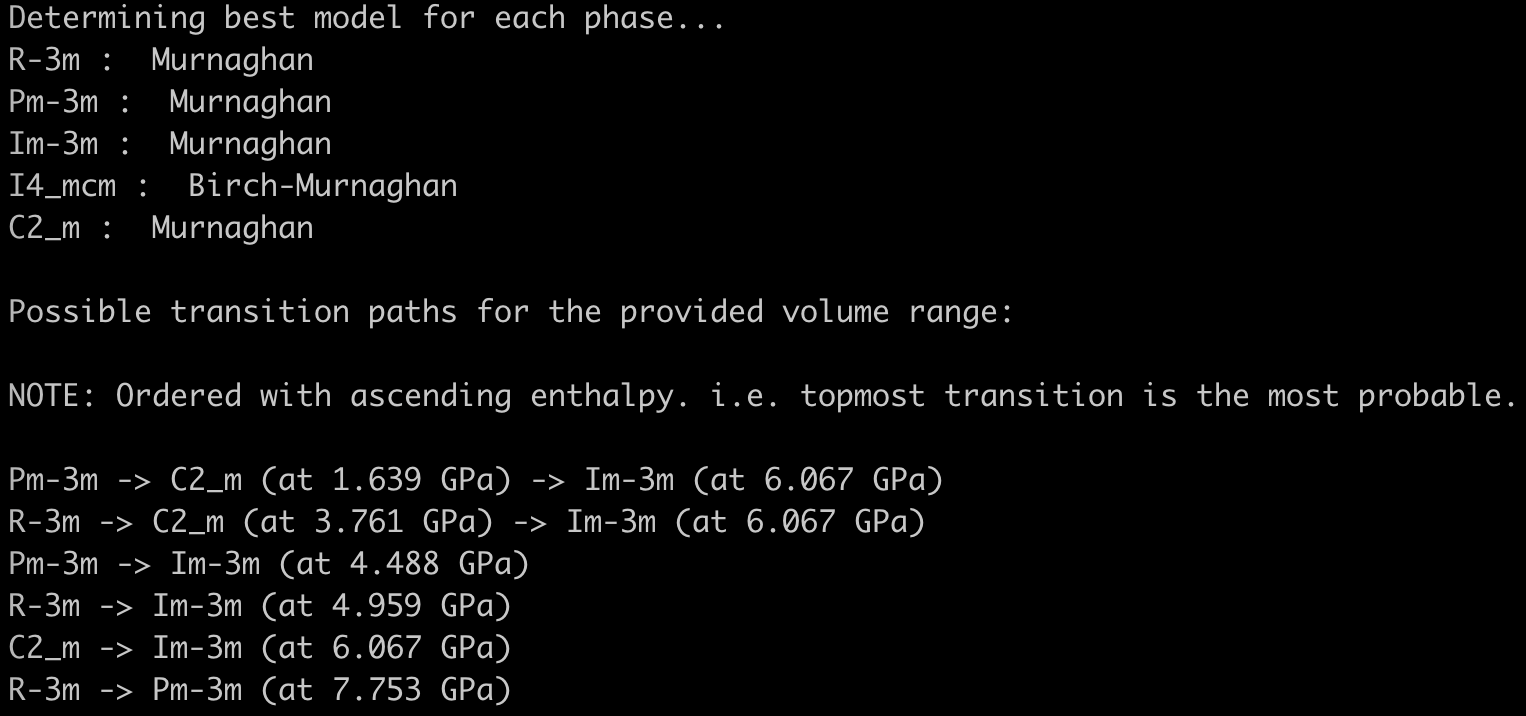}
  \caption{Possible phase transitions of Bi as calculated using the \texttt{plot\_enthalpy\_curves} function. The top most transition has the lowest enthalpy, hence it is energetically the most favorable phase transition.}
 \label{fig:phase-transitions}
     \end{center}
\end{figure}

\begin{figure}[htb!]
    \begin{center}
  \includegraphics[width=0.75\linewidth]{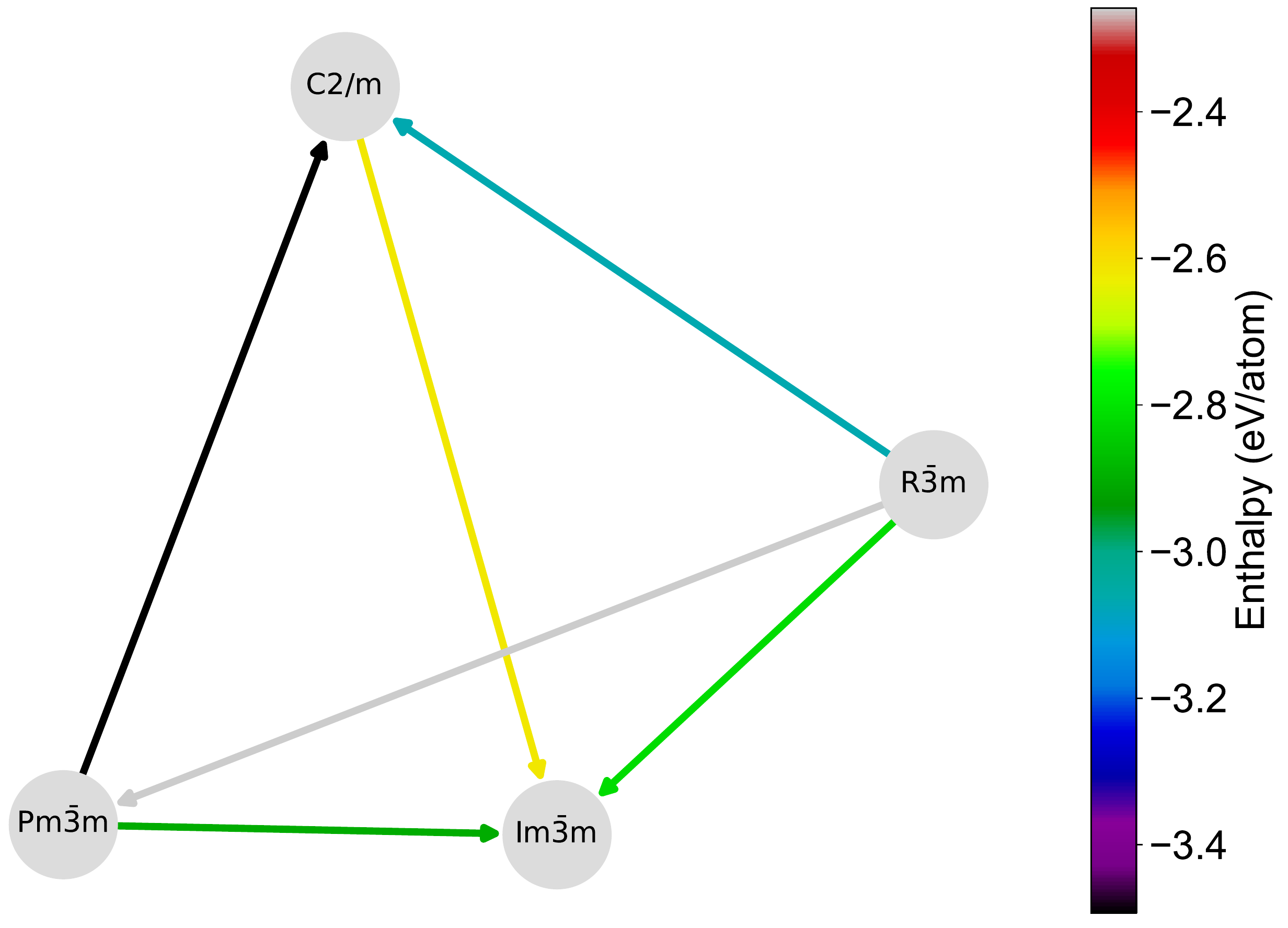}
  \caption{A network plot of the phase transition paths for different phases of Bi. The color of the arrows represents the relative enthalpy for different phase transitions. }   
 \label{fig:phase-transition-network}
     \end{center}
\end{figure}

The `Enthalpy vs. Pressure' curves for five different phases of bismuth are plotted in Figure~\ref{fig:enthalpy}. 
The data used in these calculations were obtained using the VASP code with the PBE~\cite{PBE} exchange-correlation functional. An energy cutoff of 350\,eV was employed together with k-grids of size $6 \times 6 \times 12$, $5 \times 5 \times 10$, $12 \times 12 \times 12$, $14 \times 14 \times 14$, and $10 \times 10 \times 4$ for phases $C2/m$, $I4/mcm$, $Im\bar{3}m$, $Pm\bar{3}m$, and $R\bar{3}m$, respectively.  
Additionally, the phase-transition pressures and their corresponding enthalpies were calculated for all the intersections among the `Enthalpy vs. Pressure' curves for different phases. This is particularly helpful in identifying phase boundaries and the potential phase transitions. The output of this calculation is shown in Figure~\ref{fig:phase-transitions}. The transition paths are ordered with ascending enthalpy. This means that the first transition is energetically the most favorable one and the last transition is the least favorable.
This is further elucidated in Figure~\ref{fig:phase-transition-network}, where a network plot is presented with a color bar depicting the relative energetics of different phase transition paths. 
Furthermore, the enthalpy differences with respect to a certain phase can be plot with \texttt{deltaH\_index=<phase\_index>}, where \texttt{phase\_index} is the index of the file corresponding to the baseline phase.

\bigskip
\noindent
\textbf{Usage:}
\begin{lstlisting}[language=Python]
from mechelastic import EOS
eos_object = EOS()
infiles = [`R-3m', `Pm-3m', `Im-3m', `I4_mcm', `C2_m']
natoms = [6, 1, 2, 9, 4]
eos_object.plot_enthalpy_curves(infiles, natoms, au=False)
# set au=True to convert units from eV to Ha. 
\end{lstlisting}

\section{Library Overview}
\label{sec:algorithm}
\begin{figure}[ht!]
    \begin{center}
  \includegraphics[width=1.15\linewidth]{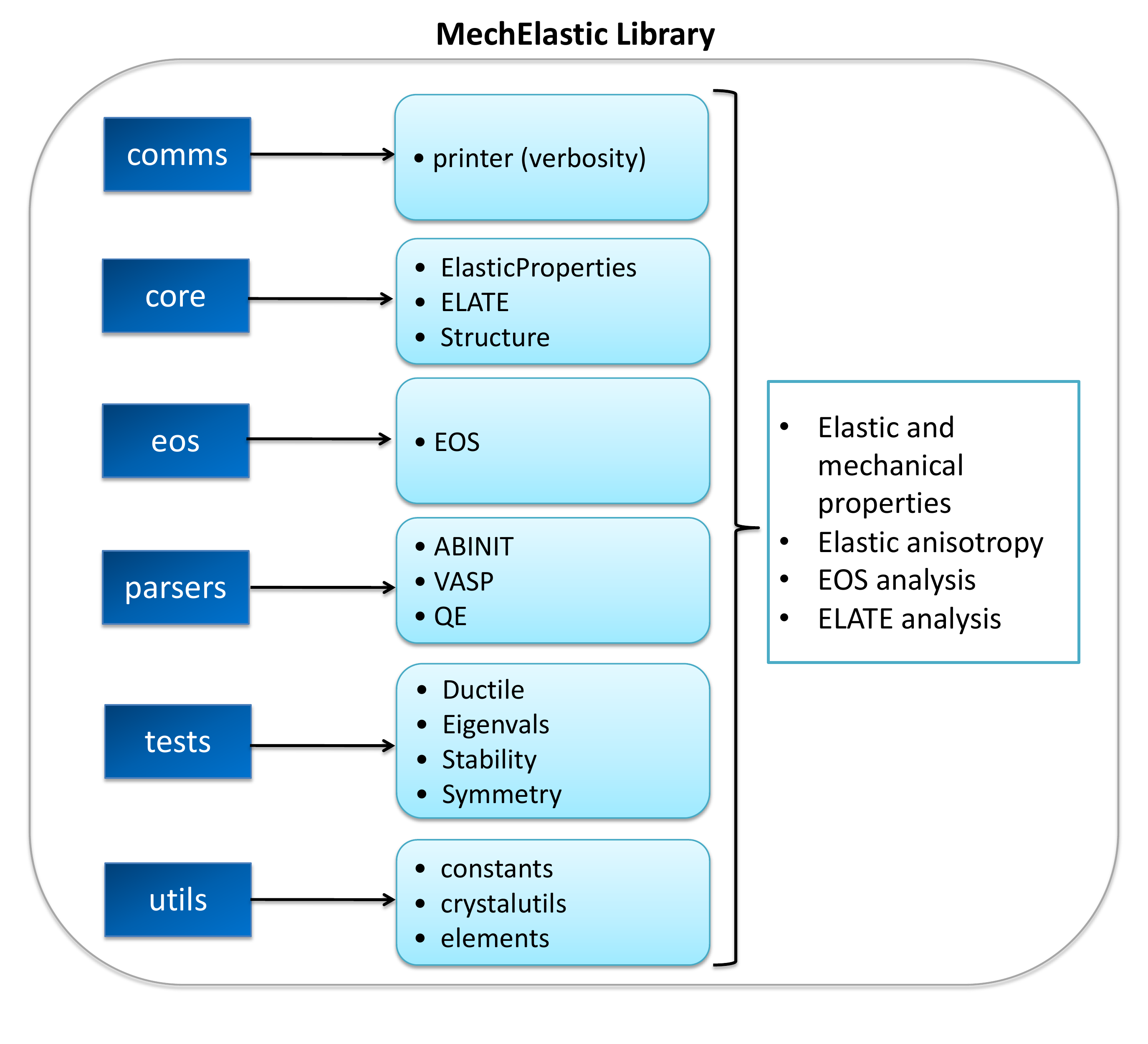}
  \caption{A structural overview of the {\sc MechElastic} library}  
 \label{fig:library}
     \end{center}
\end{figure}

{\sc MechElastic} package makes the best out of the Python's object-oriented nature, and utilizes classes and methods for the optimal efficiency. {\sc MechElastic} consists of six main components namely, \texttt{comms}, \texttt{core}, \texttt{eos}, \texttt{parsers}, \texttt{tests}, and \texttt{utils} to perform a multitude of tasks ranging from parsing the data generated by DFT calculations to testing the structure stability, EOS fittings, and providing estimation of various useful mechanical and elastic moduli. A brief overview of {\sc MechElastic} library is shown in Fig.~\ref{fig:library}. The six main  components are briefly explained below.

\begin{enumerate}

  \item \texttt{comms}: This contains methods to print messages (verbosity) to the screen such as the {\sc MechElastic} welcome screen, warning messages, and formatting of the matrices with proper precision. 
  \item \texttt{core}: This is the heart of the {\sc MechElastic} library. It contains the main code to calculate the elastic properties of 2D and 3D materials. The ELATE library is also housed here. Other than that, the \texttt{core} contains the \texttt{Structure} class of an assortment of methods to calculate structure-related properties ranging from the cell volume to the space group symmetry of the structure. 
  \item \texttt{eos}: It contains scripts associated with the EOS fittings (see Section~\ref{sec:eos}).
  \item \texttt{parsers}: This segment parses the output data from DFT calculations, mainly the elastic tensor and other quantities such as structural properties and pressure. It currently supports parsing output from the VASP, ABINIT, and Quantum Espresso codes. In the case of VASP,  whenever there is a non-zero hydrostatic pressure, the elastic tensor needs to be corrected for the residual pressure on cell. This is explained in more detail in~\ref{sec:pressure_correction}.
  \item \texttt{tests}: To  correctly study the elastic properties of materials, it is necessary to perform certain tests. This class contains tests for ductility, positive eigenvalues of the elastic tensor, and the Born-Huang mechanical stability tests for different crystal classes, as described in Section \ref{sec:mechanical}. Another test is the symmetry of the elastic tensor. 
  \item \texttt{utils}: This class contains  several handy utilities including a dictionary of commonly used constants, the atomic masses and the atomic number of elements, and a method to retrieve the crystal symmetry calculated from the crystal structure using the \texttt{spglib} library~\cite{togo2018spglib}.

\end{enumerate}

\subsection{Installation} 
The latest stable version of the {\sc MechElastic}, version 1.1.19 at the time of writing this paper, can be installed using the Python Packaging Index (\texttt{pip}) using the following command:
\begin{lstlisting}
pip install mechelastic
\end{lstlisting}
or
\begin{lstlisting}
pip3 install mechelastic
\end{lstlisting}
\noindent
The project's GitHub repository is located at\\ \url{https://github.com/romerogroup/MechElastic}. \\

\noindent An easy to follow documentation with examples can be found at\\ \url{https://romerogroup.github.io/mechelastic/}.\\

\noindent The {\sc MechElastic} package is supported by Python 3.x.

\section{Features and Implementations of the {\sc MechElastic} Package}
\label{sec:features}

\subsection{Mechanically stability test}
\begin{figure}[htb!]
\begin{center}
  \includegraphics[width=1.0\linewidth]{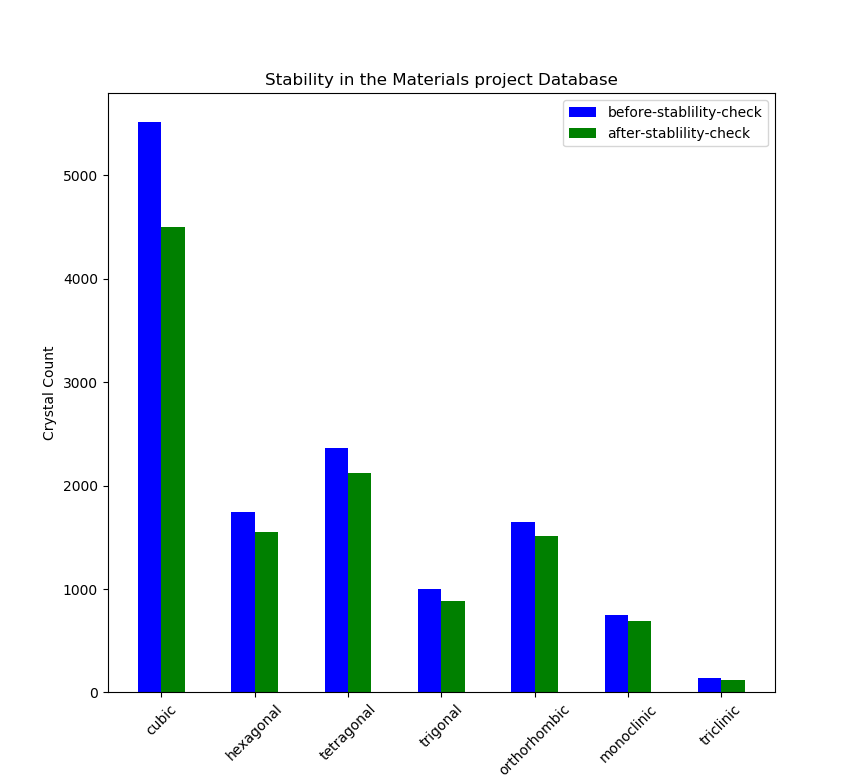}
  \caption{Use of {\sc MechElastic} in the stability screening of structures obtained from the materials project (MP) database. The total number of materials with available elastic tensor in the MP database is 13,167 (at the time of writing this paper). After the stability check, 11,390 structure are determined to be mechanically stable.}  
 \label{fig:MPstability}
\end{center}
\end{figure}

The materials project~\cite{Jain2013, AJAIN2011} database has numerous compounds reported with elastic, electronic, and thermal properties. Of those compounds, 13,167 structures have been listed with an elastic tensor. This readily attainable elastic and structure information is useful for determining trends in properties with relation to their structures. To begin this analysis, {\sc MechElastic} is used as a screening tool to filter out the mechanically unstable structures by checking for positive eigenvalues and the Born-Huang mechanical stability criteria. The results have been filtered by crystal system and are shown in Fig.~\ref{fig:MPstability}. 
Of the total 13,167 structures collected from the materials project database, 11,390 structures are found to be mechanically stable at the DFT level.

\begin{figure}[htb!]
    \center
        \includegraphics[width=1.0\textwidth]{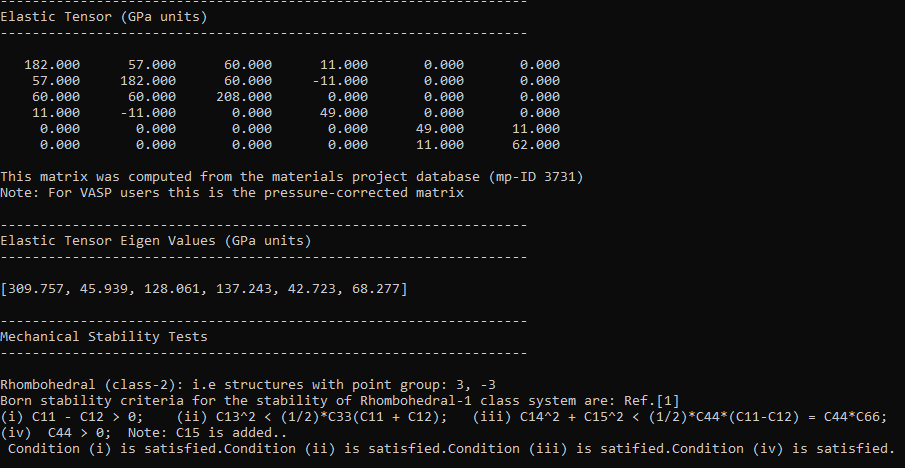}
    \caption{
            {\sc MechElastic}'s \texttt{ print\_properties()} for stability check. 
            }
    \label{fig:stability_summary}
\end{figure}

\begin{figure}[htb!]
    \center
        \includegraphics[width=1.0\textwidth]{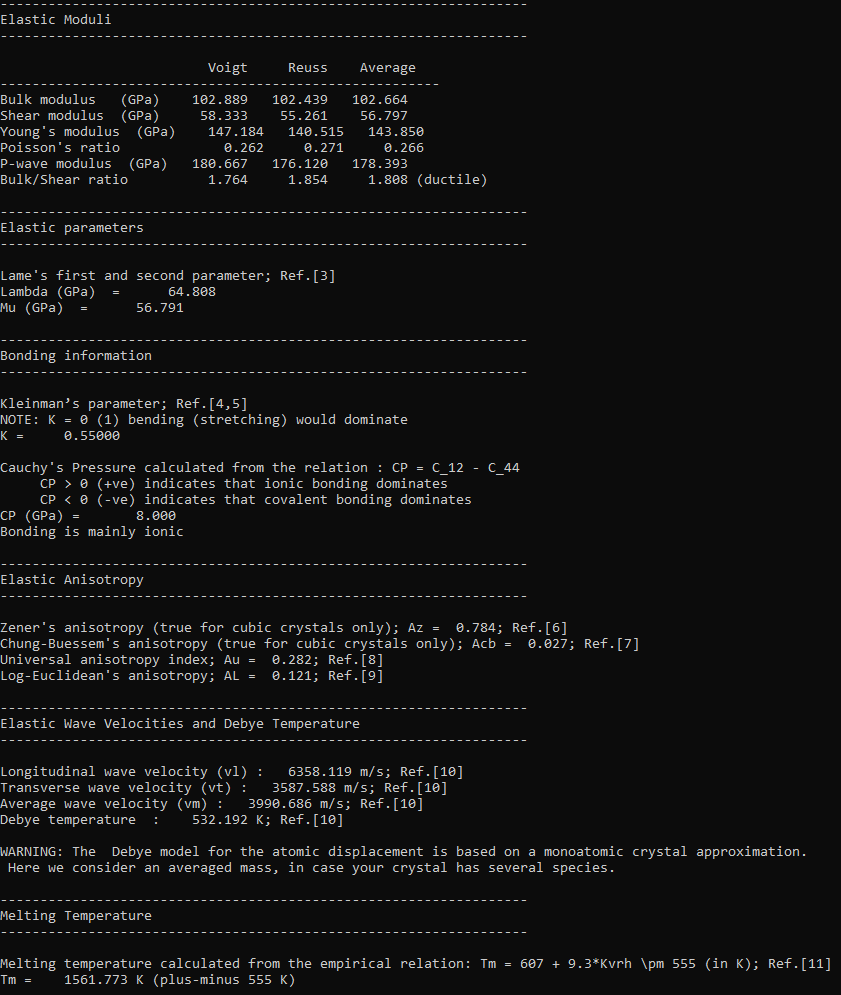}
    \caption{
            {\sc MechElastic}'s \texttt{ print\_properties()} for  elastic properties. 
            }
    \label{fig:elastic_summary}
\end{figure}

\begin{figure}[t]
    \centering
    \begin{subfigure}{0.99\textwidth}
        \includegraphics[width=\textwidth]{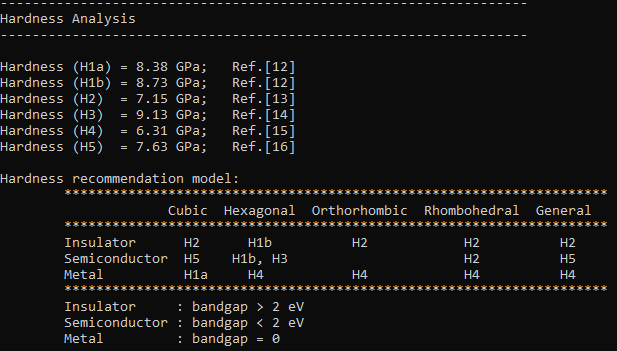}
    \end{subfigure}
    \caption{
      {\sc MechElastic}'s \texttt{ print\_properties()} for hardness anlysis.  
    }
    \label{fig:hardness_summary}
\end{figure}

\begin{figure}[htb!]
    \centering
        \includegraphics[width=\textwidth]{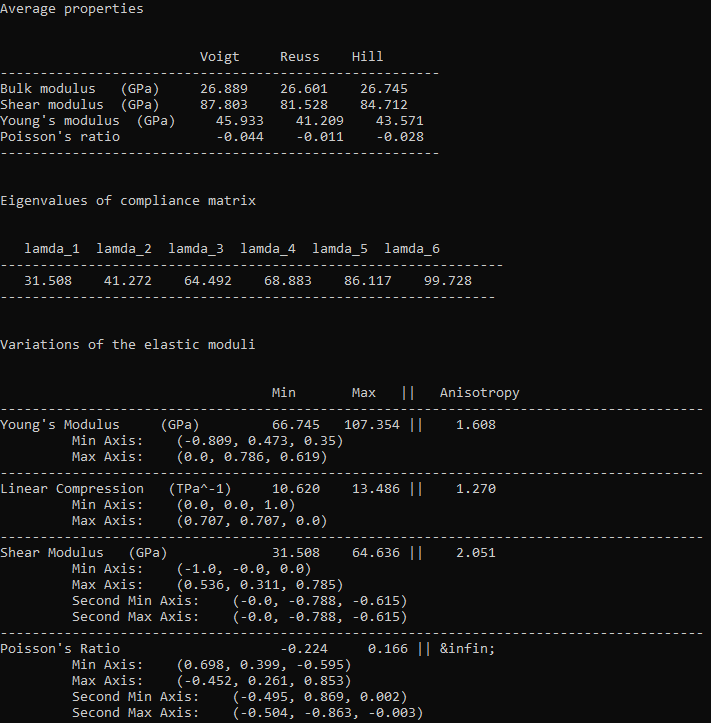}
    \caption{ELATE's output inside {\sc MechElastic}.  
    }
    \label{fig:elate_summary}
\end{figure}

\begin{figure*}[htb!]
    \centering
        \includegraphics[width=0.8\textwidth]{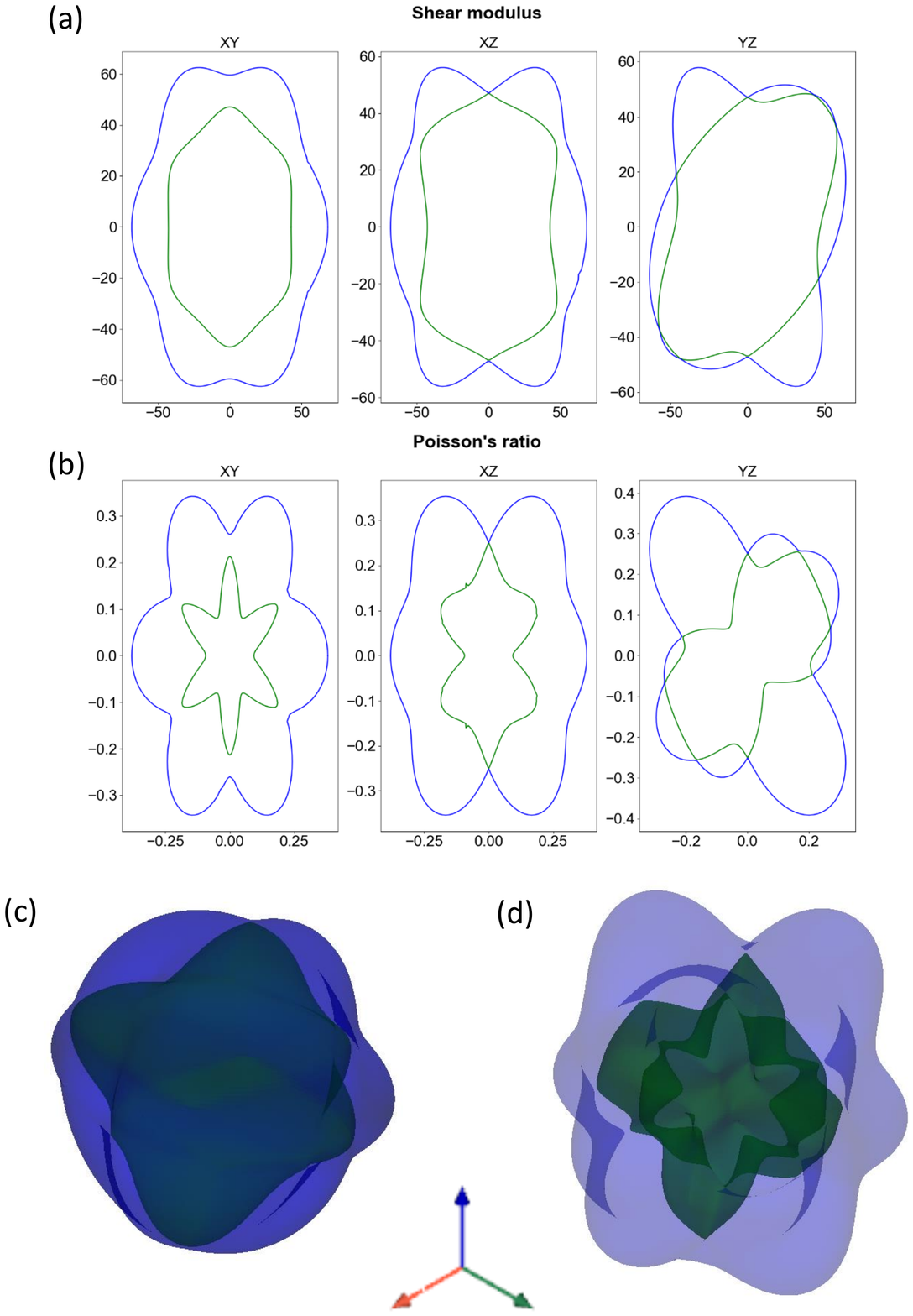}
    \caption{2D and 3D examples of the spatial dependence of the shear modulus (a, c) and Poisson's ratio (b, d) obtained by the ELATE-{\sc MechElastic} interface. The green and blue surfaces are the minimal and maximal positive value the elastic property takes, respectively. 
    }
    \label{fig:Elastic2dand3d}
\end{figure*}

\subsection{LiNbO\textsubscript{3} \sc{MechElastic}  implementation}

In this section we demonstrate the implementation of using {\sc MechElastic} on the example structure LiNbO$_{3}$. This system is chosen since it is known for having unique elastic properties, possessing piezoelectricity, and it is readily accessible on the materials project database. It's \texttt{mp-ID} is \texttt{mp-3731}, and this specific example can be readily accessed with the \texttt{mechelastic\_w\_mpDatabase.py} script in the \texttt{examples} folder of the {\sc MechElastic} package. This script demonstrates how to properly use {\sc MechElastic} to access properties in an object-oriented fashion.

The main summary of the {\sc MechElastic} package's output screen is displayed in Figs.~\ref{fig:stability_summary} - \ref{fig:hardness_summary}. It gives a concise overview of the estimated properties. Figure \ref{fig:stability_summary} shows the summary's mechanical stability check through the Born-Huang criteria and  the positive eigenvalues check of the compliance tensor. After checking the mechanical stability, the user can move on to analyze the elastic properties and other estimates that {\sc MechElastic} provides, as shown in Fig.~\ref{fig:elastic_summary}. 
The last part of the {\sc MechElastic} output prints the hardness analysis and the best hardness recommendation model. All these sections are printed subsequently, followed by a list of references so that users can easily find the paper which contains the relevant information on certain elastic estimates.

The print summary of the {\sc MechElastic}-ELATE interface is shown in Fig.~\ref{fig:elate_summary}. 
We have kept the to the same print format as done by the online ELATE interface, so that users can have the same familiarity. The {\sc MechElastic}-ELATE interface outputs a table showing the spatial average of elastic moduli computed using different averaging schemes, eigenvalues of the compliance matrix to show the stability of the structure, and information about the spatial variation of various elastic moduli. We have also kept the ploting format as true to the original as possible. A demonstration of the plots can be seen in Fig.~\ref{fig:Elastic2dand3d}, where we use the Poisson's ratio and shear modulus as an example. Figures~\ref{fig:Elastic2dand3d}(a) and  \ref{fig:Elastic2dand3d}(b) show the XY, XZ, and YZ 2D cross-section plots of the shear modulus and Poisson's ratio, respectively. Figures~\ref{fig:Elastic2dand3d}(c) and  \ref{fig:Elastic2dand3d}(d) show a snapshot of the 3D distribution of the shear modulus and Poisson's ratio, respectively. In {\sc MechElastic}, we use \texttt{pyvista's} GUI to make the plots. This gives users the ability to rotate the structures and gain a perception of the elastic anisotropy in the  studied material.

\section{Summary}
\label{sec:summary}
In summary, {\sc MechElastic} is a user friendly open-source Python library that offers various tools to carry out analysis of the elastic and mechanical properties of bulk as well as of 2D materials. 
It also provides an offline interface to the ELATE software~\cite{elate, elate2016} facilitating the study of three-dimensional spatial dependence of Young's modulus, linear compressibility, shear modulus, and Poisson's ratio. 
The current version of {\sc MechElastic} parses the elastic tensor data generated from the VASP, ABINIT, and Quantum Espresso  packages (in future we plan to support others DFT packages). 
Notably, {\sc MechElastic} assists the users (especially, the apprentices) to correctly study the elastic properties of materials by doing following: 
(i) By making sure that the elastic tensor under investigation represents a material system that is mechanically stable, $i.e.$, Born-Huang  conditions are satisfied, 
(ii) It ensures that the elastic tensor is properly corrected in case of any residual pressure on the unit cell (for VASP users), and 
(iii) It provides various warning messages and comments along with proper references to help the user in better understanding the underlying physics of the system. 
{\sc MechElastic} is specifically designed to be easily integrated in high throughput calculations. 
From the inputted elastic tensor data, the current version of {\sc MechElastic} can compute or estimate (using empirical relations) the following quantities: bulk, shear, and Young's elastic moduli, Poisson's ratio, Pugh's ratio, P-wave modulus, longitudinal  and  transverse  elastic wave velocities, Debye temperature, elastic anisotropy, 2D layer modulus, hardness, Cauchy's pressure, Kleinman's parameter, and Lame's coefficients.
Furthermore, one can analyze several equation of state models such as Vinet, Murnaghan, Birch, and  Birch-Murnaghan for a given two column `Pressure {\it versus} Volume' or `Energy {\it versus} Volume' data, and further calculate the transition pressures if datasets for multiple phases are present.


\appendix
\section{Pressure dependence of elastic constants in VASP}
\label{sec:pressure_correction}

Unlike in several other DFT codes such as ABINIT, in VASP the elastic tensor components need to be adjusted for any residual pressure on the cell while performing elastic constants analysis for the case of non-zero hydrostatic pressure~\cite{MouhatPRB2014, MOSYAGIN201720}. 

To overcome this issue, hydrostatic pressure (P) is subtracted from the diagonal components (C$_{ii}$) and added to C$_{12}$, C$_{13}$, C$_{23}$, C$_{21}$, C$_{31}$ and C$_{32}$. This is elaborated in the following matrix representation. 
\bigskip

\begin{equation}
\begin{bmatrix}
 C_{11}-P   &   C_{12}+P  &    C_{13}+P    &  C_{14}  &    C_{15}  &  C_{16} \\
 C_{21}+P  &    C_{22}-P  &    C_{23}+P   &   C_{24}  &    C_{25}  &   C_{26} \\
 C_{31}+P  &    C_{32}+P   &   C_{33}-P    &  C_{34}  &    C_{35}  &  C_{36} \\
 C_{41}  &    C_{42}  &   C_{43}   &   C_{44}-P  &   C_{45}  &   C_{46} \\
 C_{51}  &    C_{52}   &   C_{53}  &    C_{54}  &  C_{55}-P  &   C_{56} \\
 C_{61}   &   C_{62}  &    C_{63}   &   C_{64}  &    C_{65}  &  C_{66}-P 
\end{bmatrix}
\end{equation}
\bigskip

By default the pressure correction is enabled if {\sc MechElastic} detects a non-zero pressure in VASP's output. However, it could be disabled with the flag  
\texttt{adjust\_pressure=False}.

\bigskip
\noindent \emph{\bf Acknowledgements:}
We acknowledge the support from the XSEDE facilities which are supported by the National Science Foundation under grant number ACI-1053575. 
The authors also acknowledge 
the support from the Texas Advanced Computer Center (Stampede2) and  the Pittsburgh supercomputer center (Bridges), and the support from the Super Computing System (Thorny Flat) at WVU that is funded in part by the  National Science Foundation (NSF) Major Research Instrumentation Program (MRI) Award \#1726534.
AHR acknowledges the support of DMREF-NSF 1434897, NSF OAC-1740111 and DOE DE-SC0021375 projects.

\bibliographystyle{elsarticle-num_SS}
\bibliography{mechelastic.bib,references-zotero-uthpala.bib}

\end{document}